%% file: condef-main.tex
\documentclass[11pt]{elsart}
\usepackage{graphics,color,epsfig,amsmath,amssymb}
\usepackage{feynmp}
\newcounter{bla}

\def\be{\begin{align}}
\def\ee{\end{align}}
\def\bea{\begin{align}}
\def\eea{\end{align}}
\def\nn{\nonumber}

\newcommand{\secdec}{{\textsc{SecDec}}}
\newcommand{\idel}{i\,\delta}
\newcommand{\eps}{\epsilon}
\newcommand{\rd}{{\mathrm{d}}}

\begin{document}

\begin{frontmatter}
\hfill{IPPP/12/22, DCPT/12/44, MPP-2012-75}\\ 

\title{Numerical evaluation of multi-loop integrals for arbitrary kinematics with SecDec 2.0}

\author[a]{Sophia Borowka},
\author[b]{Jonathon Carter},
\author[a]{Gudrun Heinrich}

\address[a]{Max-Planck-Institute for Physics, F\"ohringer Ring 6, 80805 M\"unchen, Germany}
\address[b]{IPPP, Department of Physics, University of Durham, Durham DH1 3LE, UK}

\begin{abstract}

We present the program SecDec 2.0 which contains various new features:
First, it allows the numerical evaluation of multi-loop integrals 
with no restriction on the kinematics.
Dimensionally regulated ultraviolet and infrared singularities are isolated via 
sector decomposition, while threshold singularities are handled by a deformation of 
the integration contour in the complex plane.
As an application we present numerical results for various massive two-loop 
four-point diagrams.
SecDec 2.0 also contains new useful features for the calculation of more general 
parameter integrals, related e.g. to phase space integrals.

\begin{flushleft}
\end{flushleft}


\end{abstract}

\end{frontmatter}

{\bf PROGRAM SUMMARY}

\begin{small}
\noindent
{\em Numerical evaluation of multi-loop integrals for arbitrary kinematics with SecDec 2.0}                                       \\
{\em Authors: S.~Borowka, J.~Carter, G.~Heinrich}                                                \\
{\em Program Title: SecDec 2.0}                                          \\
{\em Journal Reference:}                                      \\
{\em Catalogue identifier:}                                   \\
{\em Licensing provisions: none}                                   \\
{\em Programming language: Wolfram Mathematica, perl, Fortran/C++}        \\
{\em Computer: from a single PC to a cluster, depending on the problem}  \\
{\em Operating system: Unix, Linux}                                       \\
{\em RAM:} depending on the complexity of the problem                                              
{\em Classification:}                                         \\
  4.4 Feynman diagrams, 
  5 Computer Algebra, 
  11.1 General, High Energy Physics and Computing.\\
{\em Journal reference of previous version:} Comput.Phys.Commun. 182 (2011) 1566-1581.                 \\

{\em Nature of problem:}\\
  Extraction of ultraviolet and infrared singularities from parametric integrals 
  appearing in higher order perturbative calculations in gauge theories. 
  Numerical integration in the presence of integrable singularities 
  (e.g. kinematic thresholds). \\
   \\
{\em Solution method:}\\
 Algebraic extraction of singularities in dimensional regularisation using iterated sector decomposition. 
 This leads to a Laurent series in the dimensional regularisation parameter $\epsilon$, 
 where the coefficients are finite integrals over the unit-hypercube. 
 Those integrals are evaluated numerically by Monte Carlo integration.
 The integrable singularities are handled by 
 choosing a suitable integration contour in the complex plane, in an automated way.
   \\
{\em Restrictions:} Depending on the complexity of the problem, limited by 
memory and CPU time.
The restriction  that multi-scale integrals could only be evaluated at Euclidean points
is superseded in version 2.0.\\
   \\
{\em Running time:}\\
Between a few minutes and several days, depending on the complexity of the problem.\\
   \\
\end{small}

\newpage


\section{Introduction}

\input{intro}

\section{General framework}
\label{sec:th}

\input{general}

\section{The \secdec{} program}
\label{sec:program}

\input{program}

\section{Installation and usage}
\label{sec:install}

\input{usage}

\section{Examples and Results}
\label{sec:examples}

\input{examples}

\section{Conclusions}

We have presented the program \secdec\,2.0, which can be used to 
factorise dimensionally regulated singularities and numerically 
calculate multi-loop integrals in an automated way. 
As a new feature of the program, it now can deal with fully physical 
kinematics, i.e. is not restricted to the Euclidean region anymore.
A new construction of the integrand,  based entirely on topological rules, 
is also included. The new features are demonstrated 
by several examples, among them a massive two-loop four-point function which is not yet known
analytically.
In addition, the program  can 
produce numerical results for more general parameter integrals, as they occur 
for example in phase space integrals for multi-particle production with several 
unresolved massless particles. The program also offers the possibility to include symbolic 
functions which can be used for instance to define measurement functions like jet algorithms 
in a flexible way. 
The program setup is such that the evaluation of several functions in parallel  
can lead to a major speed-up. To calculate full 
two-loop amplitudes involving several mass scales, the timings still leave room for improvement, 
but considering the fact that the method is very suitable for intense parallelisation, 
we are convinced that the program will be a very useful tool for a multitude of 
applications to higher order corrections in quantum field theories.

\section*{Acknowledgements}
We would like to thank Zoltan Trocsanyi, Nicolas Greiner, 
Andreas von Manteuffel  
and Pier Francesco Monni for useful comments on the program, 
and Thomas Hahn and Peter Breitenlohner for advice in computing issues.
This research was supported in parts by the British Science and Technology Facilities Council (STFC), 
and by the Research Executive Agency (REA)
of the European Union under the Grant Agreement number
PITN-GA-2010-264564 (LHCPhenoNet).

\appendix
\renewcommand \thesection{\Alph{section}}
\renewcommand{\theequation}{\Alph{section}.\arabic{equation}}
\setcounter{equation}{0}

\section{User manual}
\input{appendix}
\label{sec:appendix}
\bibliography{contour}

\end{document}

%% file: intro.tex
Currently we are in the fortunate situation of being confronted with a 
wealth of high energy collider physics data, enabling us to test our present 
understanding of fundamental interactions and to
explore physics at the TeV scale.
However, the   accuracy which has been or will be reached by the experiments 
has to be matched by comparable precision in the theory predictions, 
and in most cases this means that calculations 
beyond the leading order in perturbation theory are necessary. 

It is well known that in the calculation of higher order corrections, 
various types of singularities can arise at intermediate stages of the calculation.
For example, loop integrals can contain ultraviolet (UV) as well as infrared (IR) singularites, 
phase space integrals over unresolved massless particles  lead to infrared singularities, 
and there can be integrable singularities due to kinematic thresholds.
The UV and IR singularities can be regularised by dimensional regularisation, 
such that they appear as poles in $1/\eps$, which cancel 
when the different parts of the calculation are combined to a physical observable.
However, before such cancellations are possible, the $1/\eps$ poles have to be extracted.  
In the calculation of multi-loop integrals or real radiation  at higher orders, this 
usually leads to the task of factorising the poles from 
complicated multi-parameter integrals. 
Sector decomposition\,\cite{Binoth:2000ps,Roth:1996pd,Hepp:1966eg} is a method 
to achieve such a factorisation.
The program \secdec{}\,1.0, presented in \cite{Carter:2010hi}, performs this task 
in an automated way.
Other public implementations of sector decomposition can be 
found in \cite{Bogner:2007cr,Smirnov:2008py,Smirnov:2009pb,Gluza:2010rn}, see also \cite{Ueda:2009xx}. 
The method already has been applied in various calculations, 
listing all of them is beyond the scope of this paper, 
for a review see\,\cite{Heinrich:2008si}.
Here we just mention that there are also fruitful combinations of sector decomposition 
with other techniques,  
e.g. non-linear transformations \cite{Anastasiou:2010pw}, 
Mellin-Barnes and 
differential equation techniques\,\cite{Smirnov:2009pb,Pilipp:2008ef,Czakon:2004wm}, 
high-energy expansions\,\cite{Denner:2004iz,Denner:2008yn}, or in the context of 
subtraction for unresolved double real radiation at
NNLO\,\cite{Czakon:2010td,Czakon:2011ve,Boughezal:2011jf,Bolzoni:2010bt,Heinrich:2002rc,Anastasiou:2003gr,GehrmannDeRidder:2003bm,Binoth:2004jv,Anastasiou:2004qd}.
A method developed over many years\,\cite{Passarino:2001jd,Ferroglia:2002mz,Ferroglia:2003yj,Actis:2004bp,Passarino:2006gv,Actis:2008ts} to calculate one- and two-loop integrals numerically in the physical region 
also partly uses sector decomposition, in combination with  
a careful analysis of the singularity structure 
of certain classes of integrals 
and the use of functional relations between loop integrands.

A limitation of the program \secdec{}\,1.0 was the fact that the numerical integration of multi-scale integrals 
was only possible for Euclidean points, or, more precisely, values of the Mandelstam invariants and masses for which the 
denominator of the integrand is guaranteed to be of definite sign. 
For physical applications which go beyond one-scale problems, 
it is however crucial to be able to deal with 
integrable singularities, usually related  to kinematic thresholds, 
in addition to the singularities in $\eps$.
The program \secdec{}\,2.0 is able to achieve this task, by an automated deformation of the 
integration contour into the complex plane.
This allows the numerical calculation of multi-scale integrals in the physical region 
in an automated way. Non-planarity of the considered integral does not add any extra complications.
Adding more mass scales also does not necessarily increase the complexity of the calculation with this method, as additional masses usually lead to a simpler IR singularity structure and therefore lead to less functions in the iterated decomposition.
Therefore one can for instance obtain numerical results for two-loop integrals 
involving several mass scales where analytic methods reach their limit.

The method of contour deformation in a multi-dimensional parameter space in the context of 
perturbative calculations has been pioneered in\,\cite{Soper:1999xk} and later has been refined in various ways
to be applied to calculations at one loop\,\cite{Binoth:2002xh,Nagy:2006xy,Binoth:2005ff,Gong:2008ww,Lazopoulos:2007ix,Lazopoulos:2008de,Becker:2010ng,Becker:2011vg} 
and at two loops\,\cite{Kurihara:2005ja,Anastasiou:2007qb,Anastasiou:2008rm,Beerli:2008zz}.\\
Another purely numerical method uses an extrapolation from large to small values 
of the (analytically infinitesimal)  parameter moving the 
integration contour away from poles on the real axis\,\cite{deDoncker:2004fb,Yuasa:2011ff}.
Numerical methods using dispersion relations, differential equations and/or 
numerical integration of Mellin-Barnes representations  also 
have been worked out, see e.g.\,\cite{Bauberger:1994by,Fujimoto:1995ev,Caffo:2002ch,Pozzorini:2005ff,Czakon:2005rk,Czakon:2008zk,Freitas:2010nx}.
However, most  numerical methods to calculate multi-scale integrals
beyond one loop so far are either limited to specific types of integrals, or 
the parameters for the 
numerical integration have been carefully  adapted to the individual integrals   by the authors. 

The aim of the  work presented here is to provide a public program where the user can calculate multi-scale integrals
without worrying too much about the details of the integrand. 
The singularity structure does not have to be known beforehand  
(but certainly the user has to make sure that, after the extraction of the poles regulated by dimensional regularisation, 
only integrable singularities remain).
The program contains a sophisticated procedure 
to check and adjust the contour deformation parameters to optimize the convergence. 
For complicated integrals, the convergence can nonetheless depend critically on the settings for the numerical integration;
therefore we also offer the possibility for the user to choose various parameters at the input level to tune the deformation. \\
We should note that, even though 
the functions which are produced after the factorisation of the singularites in $\eps$ are available in algebraic form, 
they are usually too complicated to be integrated analytically.
Therefore the final integration is done by the Monte Carlo methods, meaning that the precision which can be achieved is limited,
but in favour of a gain in general applicability.
We also should remark that the method is applicable to any number of loops in principle, 
however memory problems in the algebraic part where the functions are generated, 
or bad numerical convergence can be expected if the complexity of the integral is very high.
 
The structure of this paper is as follows.  In Section \ref{sec:th}, we briefly describe the general framework. 
Section \ref{sec:program} gives an overview of the structure of the program and the new features introduced in 
\secdec{} version 2.0. Section \ref{sec:install} contains installation and usage instructions, while 
examples and results are presented in Section \ref{sec:examples}. A brief user manual is given in the Appendix. 
Detailed documentation is also coming with the code which is available at {\tt http://secdec.hepforge.org}.

%% file: general.tex

The procedure of factorising endpoint singularities from parameter integrals by iterated sector decomposition is 
described in \cite{Binoth:2000ps,Carter:2010hi}.
Here our main concern is the numerical integration for physical kinematics {\it after} 
the endpoint singularities have been extracted. 
Our method to do so is based on 
contour deformation, described in detail in Section\,\ref{sec:contour}. The following section 
serves to introduce some basic concepts.

\subsection{Feynman integrals}
We choose a scalar integral  for ease of notation. Tensor integrals 
only lead to an additional function of the Feynman parameters and invariants in the numerator. 
For more details we refer to \cite{Carter:2010hi,Heinrich:2008si}.

A scalar Feynman integral $G$ in $D$ dimensions 
at $L$ loops with  $N$ propagators, where 
the propagators can have arbitrary, not necessarily integer powers $\nu_j$,  
has the following representation in momentum space:
\begin{eqnarray}\label{eq0}
G&=&\int\prod\limits_{l=1}^{L} \rd^D\kappa_l\;
\frac{1}
{\prod\limits_{j=1}^{N} P_{j}^{\nu_j}(\{k\},\{p\},m_j^2)}\nn\\
\rd^D\kappa_l&=&\frac{\mu^{4-D}}{i\pi^{\frac{D}{2}}}\,\rd^D k_l\;,\;
P_j(\{k\},\{p\},m_j^2)=q_j^2-m_j^2+i\delta\;,
\end{eqnarray}
where the $q_j$ are linear combinations of external momenta $p_i$ and loop momenta $k_l$.
Introducing Feynman parameters leads to
\begin{eqnarray}
G&=&  
\frac{\Gamma(N_\nu)}{\prod_{j=1}^{N}\Gamma(\nu_j)}
\int_0^\infty \,\prod\limits_{j=1}^{N}\rd x_j\,\,x_j^{\nu_j-1}\, 
\delta\big(1-\sum_{i=1}^N x_i\big)
\int \rd^D\kappa_1\ldots\rd^D\kappa_L\nn\\
&&\left[ 
       \sum\limits_{i,j=1}^{L} k_i^{\rm{T}}\, M_{ij}\, k_j  - 
       2\sum\limits_{j=1}^{L} k_j^{\rm{T}}\cdot Q_j +J +\idel
                             \right]^{-N\nu}\nn\\
&&\nn\\
&=&\frac{(-1)^{N_{\nu}}}{\prod_{j=1}^{N}\Gamma(\nu_j)}\Gamma(N_{\nu}-LD/2)\int
\limits_{0}^{\infty} 
\,\prod\limits_{j=1}^{N}dx_j\,\,x_j^{\nu_j-1}\,\delta(1-\sum_{l=1}^N x_l)
\frac{{\cal U}^{N_{\nu}-(L+1) D/2}}
{{\cal F}^{N_\nu-L D/2}}\nn\\
 &&\nonumber\\
&&\mbox{where}  \nonumber\\
{\cal F}(\vec x) &=& \det (M) 
\left[ \sum\limits_{j,l=1}^{L} Q_j \, M^{-1}_{jl}\, Q_l
-J -\idel\right]\label{DEF:F}\\
{\cal U}(\vec x) &=& \det (M),  \; N_\nu=\sum_{j=1}^N\nu_j\;.\nn
\end{eqnarray}


The functions ${\cal U}$ and ${\cal F}$ also can be constructed
from the topology of the corresponding 
Feynman graph~\cite{Tarasov:1996br,Smirnov:2006ry,Heinrich:2008si}, 
and the implementation of this construction in \secdec{}\,2.0
is one of the new features of the program. 

For a diagram with massless propagators, 
none of the Feynman parameters occurs quadratically in 
the function ${\cal F}={\cal F}_0$ . If massive internal lines are present, 
${\cal F}$ gets an additional term 
${\cal F}(\vec x) =  {\cal F}_0(\vec x) + {\cal U}(\vec x) \sum\limits_{j=1}^{N} x_j m_j^2$. 

${\cal U}$ is a positive semi-definite function. 
A vanishing ${\cal U}$ function is related to the  UV subdivergences of the graph. 
Overall UV divergences, if present,
will always be contained in the  prefactor $\Gamma(N_{\nu}-m-LD/2)$. 
In the region where all invariants formed from external momenta are negative, 
which we will call the {\em Euclidean region} in the following, 
${\cal F}$ is also a positive semi-definite function 
of the Feynman parameters $x_j$.  Its vanishing does not necessarily lead to 
an IR singularity. Only if some of the invariants are zero, 
for example if some of the external momenta
are light-like, the vanishing of  ${\cal F}$  may induce an IR divergence.
Thus it depends on the {\em kinematics}
and not only on the topology (like in the UV case) 
whether a zero of ${\cal F}$ leads to a divergence or not. 
The necessary (but not sufficient) conditions for a divergence 
are given by the Landau equations~\cite{Landau:1959fi,ELOP,Nakanishi}:
\begin{eqnarray}
&& x_j\,(q_j^2-m_j^2)=0 \quad \forall \,j \label{landau1}\\
&&\frac{\partial}{\partial k^\mu}\sum_j x_j\,\left(q_j^2(k,p)-m_j^2\right)=0\;.\label{landau2}
\end{eqnarray}
If all kinematic invariants formed by external momenta are negative, 
the necessary condition ${\cal F}=0$ for an IR divergence can only 
be fulfilled if some of the parameters $x_i$ go to zero.
These endpoint singularities can be regulated by dimensional regularisation and 
factored out of the function ${\cal F}$ using sector decomposition. 
The same holds for dimensionally regulated UV singularities contained in ${\cal U}$.
However, after the UV and IR  singularities have been extracted 
as poles in  1/$\epsilon$,
for non-Euclidean kinematics 
we are still faced with integrable singularities related to kinematic thresholds. 
How we deal with these singularities will be described in the following section.

\subsection{Deformation of the integration contour}
\label{sec:contour}
\begin{figure}[h]
	\begin{center}
	      \includegraphics[width=0.5\textwidth]{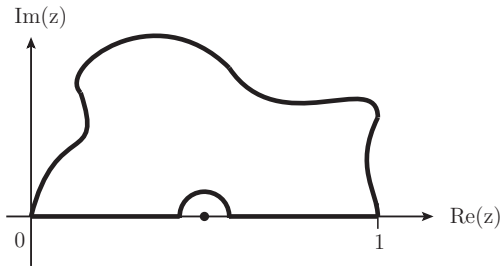}
	\end{center}
	\caption{Schematic picture of the closed contour avoiding poles on the real axis.}
	\label{fig:scenario1} 
\end{figure}
Unless the function ${\cal F}$ in eq.~(\ref{DEF:F}) is of definite sign for
all possible values of invariants and Feynman parameters, 
the  denominator of a multi-loop integral will  vanish within the integration 
region on a hypersurface given by the solutions of the Landau equations. 
If eq.~(\ref{landau1}) has a solution where $x_j > 0\, \forall \,j$, all particles in the
loop
go simultaneously on-shell. This corresponds to a leading Landau singularity, which is not
integrable (for real values of masses and momenta).
However, the integrand can also diverge for certain values of kinematical invariants and Feynman parameters
which represent a subleading Landau singularity, 
corresponding to an integrable singularity of logarithmic or squareroot type, related 
to normal thresholds.
In these cases, we can make use of Cauchy's theorem to
avoid the non-physical poles on the real axis by a deformation of the integration contour 
into the complex plane. 
As long as the deformation is in accordance with the causal $i\delta$  prescription 
of the Feynman propagators, and no poles are crossed while changing the integration
path, the integration contour can be changed such that
the convergence of the numerical integration 
is assured. Using the fact that the integral over the closed contour in 
Fig.~\ref{fig:scenario1} is zero, we have 
\begin{align}
\int_0^1 \prod\limits_{j=1}^{N}\rd x_j {\cal I}(\vec{x}) = \int_0^1 \prod\limits_{j=1}^{N}\rd x_j 
\left\vert \left(\frac{\partial z_k(\vec{x})}{\partial x_l}\right)\right\vert {\cal
I}(\vec{z}(\vec{x}))\;,
\end{align}
where the $x_i$ are real, while $z_i$ are complex, describing a path parametrized by 
the variables $x_i$.
The $i\delta$ prescription for the Feynman propagators 
tells us that the contour deformation into the complex plane should be such that 
the imaginary part of ${\cal F}$ 
should always be negative. 
For real masses and Mandelstam invariants $s_{ij}$, the following Ansatz\,\cite{Soper:1999xk,Nagy:2006xy,Binoth:2005ff}
is therefore convenient:
\begin{align}
\label{eq:condef}
\vec{z}( \vec x) &= \vec{x} - i\;  \vec{\tau}(\vec{x})\nonumber\\
\tau_k &= \lambda\, x_k (1-x_k)
 \, \frac{\partial {\cal F}(\vec{x})}{\partial x_k}  \;.
\end{align}
The derivative of ${\cal F}$ in eq.~(\ref{eq:condef}) is smallest in the extrema and largest where 
the slope is maximal. 
Hence, unless we are faced with a leading Landau singularity where both ${\cal F}$ and its
derivatives with respect to $x_i$ vanish, 
the deformation leads to a well behaved integral at the points where the function ${\cal F}$ vanishes. 
A closed integration contour is guaranteed by the factors $x_k$ and $(1-x_k)$, keeping the endpoints  fixed. 
In terms of the new variables, we thus obtain
\begin{equation}
\label{eq:newF}
{\cal F}(\vec{z}(\vec{x}))={\cal F}(\vec{x})
-i\,\lambda\,\sum\limits_{j} \, x_j (1-x_j)\, \left(\frac{\partial {\cal F}}{\partial x_j}  \right)^2 + {\cal O}(\lambda^2)\;,
\end{equation}
such that ${\cal F}$ acquires a negative imaginary part of order $\lambda$. Hence, the size of $\lambda$ determines the scale of the deformation.
More technical details about the deformation are given in Section \ref{sec:program:contourdef}.

\subsection{Parameter integrals}

The program \secdec{} can also factorise singularities from parameter integrals which are  more 
general than the ones related to multi-loop integrals.
The only restrictions are that the integration domain should be the unit hypercube, 
and the singularities should be only endpoint singularities, i.e. should be located at zero 
or one. Contour deformation is not available in the subdirectory {\tt general}, because 
the sign of the imaginary part telling us how to deform the contour is not fixed a priori for general functions, in contrast
to  loop integrals. However, we plan to implement the use of Cauchy's theorem where applicable 
in a future version.
Currently we assume that the singularities are regulated 
by non-integer powers of the integration parameters, where the non-integer part is the 
$\eps$ of dimensional regularisation or some other regulator.
The general form of the integrals is 
\begin{equation}
I=\int_0^1 dx_1 \ldots \int_0^1 dx_N \prod_{i=1}^m P_i(\vec{x},\{\alpha\})^{\nu_i}\;,
\label{eq:general}
\end{equation}
where $P_i(\vec{x},\{\alpha\})$ are polynomial functions of the parameters $x_j$,
which can also contain some symbolic constants $\{\alpha\}$. 
The user can leave the parameters $\{\alpha\}$ symbolic during the decomposition,
specifying numerical values only for the numerical integration step.
This way the decomposition and subtraction steps do not have to be redone if 
the values for the constants are changed.
The $\nu_i$ are powers of the form $\nu_i=a_i+b_i\epsilon$ 
(with  $a_i$ such that the integral is convergent).
Note that half integer powers are also possible.

%% file: program.tex

\subsection{Structure}

\begin{figure}[htb]
\includegraphics[width=14cm]{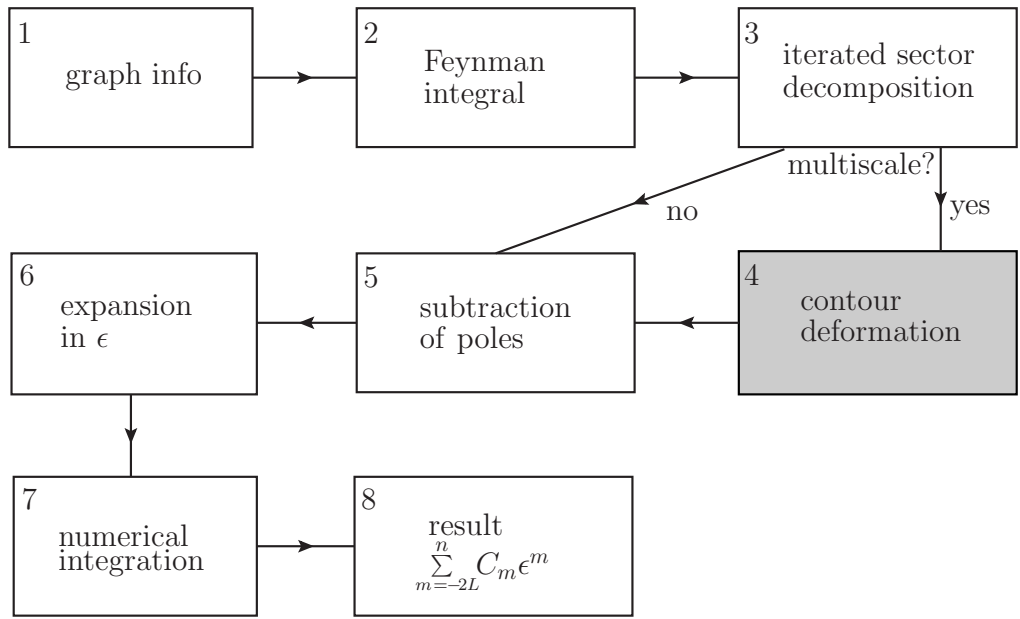}
\caption{Flowchart showing the main steps the program 
performs  to produce the result as a Laurent series in $\eps$.}
\label{fig:flowchart}
\end{figure}

The program consists of two parts, an algebraic part and a numerical part.
The algebraic part uses code written in Mathematica~\cite{Wolfram} and does the 
decomposition into sectors, the subtraction of the 
singularities, the expansion in $\eps$ and the generation of the 
files necessary for the numerical integration. 
In the numerical part,  Fortran or C++ functions forming the coefficient 
of each term in the Laurent series in $\eps$ are integrated using the 
Monte Carlo integration programs contained in the
\textsc{Cuba} library\,\cite{Hahn:2004fe,Agrawal:2011tm}, or 
\textsc{Bases}\,\cite{Kawabata:1995th}. 
The different subtasks are handled by perl scripts.  
The flowchart of the program is shown in Fig.~\ref{fig:flowchart} for the basic 
building blocks to calculate multi-loop integrals. To calculate  parameter 
integrals which are not necessarily related to loop integrals, 
the structure is the same except that contour deformation is not available.
For more details about the features in the part {\tt general} we refer to \cite{Carter:2010hi}.

The  directories {\tt loop} and {\tt general} have the same global structure, 
only some of the individual files are specific to loop integrals or to more general parametric functions.
The directories contain a number of perl scripts steering the 
decomposition and the numerical integration.  The scripts use perl modules contained 
in the subdirectory {\tt perlsrc}.

The Mathematica source files 
are located in the subdirectories {\tt src/deco} (files used for the decomposition), 
{\tt src/subexp} (files used for the pole subtraction and expansion in $\eps$) and 
{\tt src/util} (miscellaneous useful functions).  
The documentation, created by {\it robodoc}\,\cite{robodoc} 
is contained in the subdirectory {\tt doc}. It contains an index to look up documentation 
of the source code in html format by loading {\tt masterindex.html} into a browser.

\vspace*{3mm}

In order to use the program, the user only has to edit the following two files:
\begin{itemize}
\item {\tt param.input}: (text file)\\
specification of paths, type of integrand, 
order in $\eps$, output format, parameters for numerical integration, 
further options
\item {\tt Template.m}: (Mathematica syntax)
\begin{itemize}
\item for loop integrals: specification of loop momenta and propagators, 
resp. of the topology; optionally numerator, non-standard propagator powers, space-time dimensions
\item for general functions: specification of integration variables, integrand, variables to be split
\end{itemize}
\end{itemize}
The program comes with example 
input and template files in the subdirectories {\tt loop/demos} 
respectively {\tt general/demos}, described in detail in  \cite{Carter:2010hi}.

\subsection{New features of the program}

Version 2.0 of  \secdec{} contains the following new features, 
which will be described in detail in this section, while examples will be 
given in Section \ref{sec:examples}.
\begin{itemize}
\item {\bf loop part:}
\begin{itemize}
\item Multi-scale loop integrals can be evaluated without restricting the kinematics 
to the Euclidean region. This has been achieved by performing a (numerical) contour integration 
in the complex plane. The program automatically tries to find an optimal deformation 
of the integration path.
\item For scalar multi-loop integrals, the integrand can be constructed from the topological cuts of 
the diagram. 
The user only has to provide the vertices and the propagator masses, but does not have to provide the momentum flow.
\item The files for the numerical integration  of multi-scale loop 
diagrams with contour deformation are written in C++ rather than Fortran. 
For integrations in Euclidean space, both the Fortran and the C++-versions are supported.
The choice between Fortran and C++ can be made by the user in the {\tt param.input} file by choosing 
either {\it language=Cpp} (default) or {\it language=fortran} .
\item A parallelisation of the algebraic part for Mathematica versions 7 and higher is possible if several cores are available.
\item The most recent version of the \textsc{Cuba} library, \textsc{Cuba}-3.0 (beta), is added to the program and used by default. The older version \textsc{Cuba}-2.1 is still supported.
\item The rescaling of the kinematic invariants is now possible by choosing {\it rescale=1} (default is 0).   
\end{itemize}
\item {\bf general part:}
\begin{itemize}
\item The user can define additional (finite) functions at a symbolic level and specify them only later 
after the integrand has been transformed into a set of finite parameter integrals for each order in 
$\epsilon$. 
\end{itemize}
\item {\bf both parts:}
\begin{itemize}
\item The possibility to loop over ranges of parameter values is automated.
\end{itemize}
\end{itemize}
In the following we describe the new features in more detail.

\subsection{Multi-scale loop integrals: Implementation of the contour deformation}
\label{sec:program:contourdef}

As explained in Section \ref{sec:contour}, singularities on the real axis can be avoided 
by a deformation of the integration contour into the complex plane.
The overall size of the deformation is controlled by the parameter $\lambda$ defined in eq.~(\ref{eq:condef}).

The convergence of the numerical integration can be improved significantly 
by choosing an ``optimal" value for $\lambda$.
Values of $\lambda$ which are too small lead to contours which are too close 
to the poles on the real axis and therefore lead to bad convergence.
Too large values of $\lambda$ can modify the real part of the function 
to an unacceptable extent and could even change the sign of the
imaginary part if the terms of order $\lambda^3$ get larger than the 
terms linear in $\lambda$. This would lead to a wrong result.
Therefore we implemented a four-step procedure to optimize the value of $\lambda$, 
consisting of 
\begin{itemize}
\item ratio check: To make sure that the terms of order $\lambda^3$ in
  eq.~(\ref{eq:newF})  do not spoil the sign of the imaginary part, we
  evaluate the ratio  of the terms linear and cubic in  $\lambda$ for
  a quasi-randomly chosen set of sample points to determine the maximal
allowed $\lambda=\lambda_{max}$.
\item modulus check: The imaginary part is vital at the points where
  the real part of  ${\cal F}$ is vanishing. In these regions,
  the deformation should be large enough to avoid large numerical
  fluctuations due to a highly peaked integrand. Therefore we check
  the modulus of  each subsector function ${\cal F}_i$ at a number of sample points,
and pick the fraction of the value of $\lambda_{max}$ 
which maximises the minimum of the modulus of ${\cal F}_i$, 
i.e. the value of lambda which keeps ${\mathcal F}_i$ furthest from zero.
\item individual $\lambda(i,j)$ adjustments: If the values of
  $\frac{\partial {\cal F}_i}{\partial x_j}$  are very different in magnitude, it can be convenient to
  have an individual parameter $\lambda(i,j)$ for each subsector
  function ${\cal F}_i$ and each Feynman parameter $x_j$. 
\item sign check: After the above adjustments to $\lambda$ have been made, 
the sign of Im$(\mathcal F)$ is again checked for a number of sample
points. If the sign is ever positive, this value of $\lambda$ is disallowed.
\end{itemize}

The contour deformation can be switched on or off by choosing \\
{\it contourdef=True/False} 
in the input file {\tt param.input}. Obviously, the calculation takes longer if contour deformation
is done, so if the integrand is known to be positive definite, contour deformation should be switched off. 
We also should emphasize that for integrands with a complicated threshold structure, the success of the 
numerical integration can critically depend on the parameters which tune the deformation, 
and on the settings for the Monte Carlo integration. 
In order to allow the user to tune the deformation, the following parameters can be adjusted by the user 
in  the input file: 
\begin{description}
\item[lambda:]  the program takes the $\lambda$ value given in {\tt param.input}
as a starting point. If, after the program has performed the checks listed above, this $\lambda$ is found to be 
unsuitable or suboptimal, the value of $\lambda$ will be changed automatically by the program.
The default is {\it lambda=1.0}.
\item[largedefs:] If the integrand is expected to  have (integrable) endpoint singularities 
at $x_j=0$ or 1, the deformation should be large in order to move the contour away 
from the problematic region. If {\it largedefs=1} (default is 0), the 
program tries to enlarge the deformation at the endpoints.
\item[smalldefs:] If the integrand is expected to be oscillatory and hence sensitive to small changes in the 
deformation parameter $\lambda$, choosing the flag {\it smalldefs=1} (default is 0) will minimize the argument 
of each subsector function ${\cal F}_i$ by varying $\lambda(i,j)$.
\end{description}

\subsection{Topology-based construction of the integrand}
\label{subsec:cutconstruct}
As already mentioned in section \ref{sec:th}, the functions ${\cal U}$ and ${\cal F}$ can be constructed
from the topology of the corresponding 
Feynman graph~\cite{Tarasov:1996br,Smirnov:2006ry,Heinrich:2008si}, 
without the need to assign the momenta for each propagator explicitly. 
The user only has to  label the external momenta and the vertices. If an external momentum $p_i$ is  
part of a vertex, this vertex needs to carry the label $i$. The labelling of vertices containing only 
internal lines is arbitrary.
In {\tt Template.m}, the user has to specify {\it proplist} as a list of entries of the form 
$\{ms[k],\{i,j\}\}$, where $ms[k]$ is the mass squared of the propagator connecting vertex $i$ and vertex $j$.
The mass label $k$ must correspond the the $k$th entry of the list of masses given in 
{\tt param.input}. While $k$ needs to be the number labelling the masses, $ms[k]$ (with $k$ being an integer)
 can be left symbolic during the decomposition. However, if the mass is zero, one has to put 
 $\{0,\{i,j\}\}$, because this changes the singularity structure at decomposition level. 

An example is given below, more examples can be found in the mathematica template files 
{\tt templateP126.m, templateBnp6*.m, templateJapNP.m, templateggtt*.m}
in the subdirectory {\tt loop/demos}.
This feature of constructing the graph topologically is only implemented for scalar integrals so far. 
The original form of specifying the 
propagators by their momenta, as done in {\secdec}\,1.0, is still operational. 
The topology based construction is selected by defining {\it cutconstruct=1} in the input file.  

\subsection{Looping over ranges of parameters}
As the algebraic part can deal with symbolic expressions for 
the kinematic invariants or other parameters contained in the integrand, 
the decomposition and subtraction parts only need to be done once for the calculation
of many different numerical points. Therefore it is desirable to automate the calculation
of many numerical points to minimize the effort for the user. This is done using
the perl script {\tt multinumerics.pl}. The user should create a text file
{\tt multiparamfile} in {\tt myworkingdir}, and specify a number of options:
\begin{itemize}
\item {\it paramfile=myparamfile}: specify the name of the parameter file.
\item {\it pointname=myprefix}: points calculated will have the names {\it
myprefix1, myprefix2,...}
\item {\it lines}: the number of points you wish to calculate - if omitted all
points (listed in separate lines) will be calculated.
\item {\it xplot}: the number of the column 
containing the values which should be used on the x-axis of the plot (default is 1).
\end{itemize}
After these options, the numerical values of the parameters for each point to calculate
should be specified. 
In the {\tt loop} directory, the number of values 
given for $s_{ij}$, $p_{i}^2$ and  $m_i^2$ needs to be specified 
by numsij=, numpi2= and numms2=. An example can be found in  {\tt loop/demos/multiparam.input}.
The following example explains how the numerical values for each point are written down in the 
{\tt general} directory. 
If you wished to calculate three numerical points for a function 
where the symbols $a,b$ (defined as symbols in the parameter input file) should take on the values 
$(a,b)=(0.1,0.1),(0.2,-0.4),(-0.3,0.9)$ then the inputs in {\tt multiparamfile} for this
would be:\\
{\it 0.1,0.1\\
0.2,-0.4\\
-0.3,0.9}\\
Furthermore, one may wish to calculate the integrand for values of parameters at
incremental steps. This is allowed, and the syntax is as follows: Suppose you
wish to calculate each combination of $s=0.1,0.2,0.3$ and $t=0.1,0.3,0.5,0.7$.
The input for this is\\
{\it minvals=0.1,0.1\\
maxvals=0.3,0.7\\
stepvals=0.1,0.2}\\
Non-equidistant step values are also possible. For instance, to calculate every
combination of $a=0.1,0.2,0.4,\,b=0.1,0.3,0.6$ the syntax would be:\\
{\it values1=0.1,0.2,0.4\\
values2=0.1,0.3,0.6}\\
Please note that {\it values1} must appear before {\it values2} in
{\tt multiparamfile}.\\
Examples can be found in
{\tt general/demos/multiparam.input} or \\
{\tt loop/demos/multiparam.input}.
In the {\tt loop} directory, there is a perl script {\tt helpmulti.pl} which can be used to generate 
the files {\tt multiparam.input} automatically to avoid typing large sets of numerical
values.\\
In order to execute the script {\tt multinumerics.pl}, 
the Mathematica-generated functions must already
be in place. The simplest way to do this is to run the {\tt launch} script, with
{\it exeflag=1} in your parameter file. 
Then issue the command {\it `./multinumerics.pl [-d myworkingdir -p
multiparamfile]'}. In single-machine mode ({\it clusterflag=0})
all integrations will then be performed, and the results collated and output as
files in the directory specified in {\tt myparamfile}. In batch mode you will
need to run the script again, with the argument `1', to collect the results, i.e.
{\it `./multinumerics.pl 1 [-d myworkingdir -p multiparamfile]'}. The script
generates a parameter file for each numerical point calculated. To remove these
intermediate parameter files (your original {\it myparamfile} will not be
removed), issue the command {\it `./multinumerics.pl 2 [-d myworkingdir -p
multiparamfile]'}. This should only be done after the results have been
collated.

\subsection{Leaving functions implicit during the algebraic part}
\label{sec:dummy}

This feature is available in the part {\tt general} to evaluate 
general parametric functions, where it is possible to include a ``dummy"
function depending on (some of) the integration parameters, 
the actual form of the function being specified 
only later at the numerical integration stage.
There are a number of reasons why one might want to leave
functions implicit during the algebraic stage. 
For example,  squared matrix elements typically contain
large but finite functions of the phase space variables in the numerator, so
the algebraic part of the calculation will be
quicker and produce much smaller intermediate files if these functions are left
implicit. Also, one might like to use a number of measurement
functions and be able to specify or change them without having to redo the decomposition.
To use this option, the Mathematica template file 
can contain a function which is left undefined, but needs to be listed
 under the option {\it dummys} in the parameter input file. Note that
one may use more than one implicit function at a time, 
and that these functions can have any number of arguments. 
If symbolic parameters are also used, these
do not need to be arguments of the implicit function.\\
Once the template and parameter files are set up, the functions need to be defined
explicitly so that they can be used in the calculation. The
simplest way to do this is to prepare a Mathematica syntax file for each
implicit function specified, and place them in the {\it outputdir} specified in
your parameter file. Suppose you have a function named {\it dum1}, a function of
two variables, defined as $dum1(x_1,x_2)=1+x_1+x_2$. Then you should create a file
{\tt dum1.m}, and insert the lines:\\
$intvars=\{z_1,z_2\};\\dum1=1+z_1+z_2;$\\
where $z_1,z_2$ can be replaced by any variable name you wish, as long as they
are used consistently in {\tt dum1.m}. 
Notice that for every function specified in {\it dummys} in your parameter file, 
there must be a Mathematica file {\tt dummyname.m} with the correct name and syntax 
in the results directory. Once these Mathematica files are in place, 
issue the command\\ `{\it
createdummyfortran.pl [-d myworkingdir -p myparamfile]}' from the {\tt general}
directory. This generates the fortran files for the functions you defined, which
are found in the same subdirectory as the originals.\\
Of course you might prefer to write these fortran files yourself instead of
having them generated by the program. 
This is certainly possible, however we recommend
that you use this perl script to generate functions with the necessary
declarations and then edit these.\\
An example of this can be found in {\tt general/demos}, with the files \\
{\tt paramdummy.input, templatedummy.m}, and the directory {\tt /testdummy}.

%% file: usage.tex

\subsection{Installation}
The program can be downloaded from \\
{\tt http://secdec.hepforge.org}.

Unpacking the tar archive via 
{\it  tar xzvf SecDec.tar.gz} 
will create a directory called {\tt SecDec} 
with the subdirectories as described above. Then change to the {\tt SecDec} directory
and run {\it ./install}.

Prerequisites are Mathematica, version 6 or above, perl (installed by default on 
most Unix/Linux systems), a Fortran compiler (e.g. gfortran, ifort) or a C++ 
compiler if the C++ option is used.

\subsection{Usage}

\begin{enumerate}

\item Change to the subdirectory {\tt loop} or {\tt general}, depending 
on whether you would like to calculate a loop integral or a more general parameter integral.
\item Copy the files {\tt param.input} and {\tt template.m} to 
create your own parameter and template files  {\tt myparamfile.input}, {\tt mytemplatefile.m}.
\item Set the desired parameters in {\tt myparamfile.input} and specify the integrand in {\tt mytemplatefile.m}.
\item Execute the command {\it ./launch -p myparamfile.input -t mytemplatefile.m} 
in the shell.  \\
If you omit the option {\it -p myparamfile.input}, the file {\tt param.input} will be taken as default.
Likewise, if you omit the option {\it -t mytemplatefile.m}, 
the file {\tt template.m} will be taken as default.
If your files {\it myparamfile.input, mytemplatefile.m} are in a different directory, say, 
{\it myworkingdir}, 
 use the option {\bf -d myworkingdir}, i.e. the full command then looks like 
 {\it ./launch -d myworkingdir -p myparamfile.input -t mytemplatefile.m}, 
 executed from the directory {\tt SecDec/loop} or
 {\tt SecDec/general}. \\

\item Collect the results. Depending on whether you have used a single machine or 
submitted the jobs to a cluster, the following actions will be performed:
 \begin{itemize}
\item If the calculations are done sequentially on a single machine, 
    the results will be collected automatically (via {\tt results.pl} called by {\tt launch}).
    The output file will be displayed with your specified text editor.

\item If the jobs have been submitted to a cluster,    
	when all jobs have finished,  execute  the command 
	{\it ./results.pl [-d myworkingdir -p myparamfile]}. 
	This will create the files containing the final results in the {\tt graph} subdirectory
	specified in the input file.

\end{itemize}


\item After the calculation and the collection of the results is completed, 
you can use the shell command {\it ./launchclean[graph]}
to remove obsolete files.

\end{enumerate}

It should be mentioned that the code starts working first on the most complicated 
pole structure, which takes longest. 
This is because in case the jobs are sent to a cluster, it is advantageous to 
first submit the jobs which are expected to take longest.

%% file: examples.tex
\subsection{Massive two-loop integrals}

\subsubsection{A two-loop three-point function}
\begin{figure}[!htbp]
\begin{center}
\includegraphics[width=5.cm]{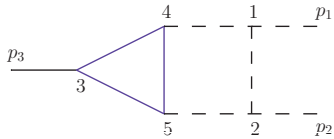}
\end{center}
\caption{Two-loop vertex graph $P_{126}$, containing a massive triangle loop.  
Solid lines are massive, dashed lines are massless. The vertices are labeled to match the 
construction of the integrand from the topology as explained in the text.}
\label{fig:P126}
\end{figure}

In this example, we will demonstrate three of the new features 
of the \secdec{} 2.0 program: the construction of  $\mathcal F, \mathcal U$  directly from the
topology of the graph,  the evaluation of the graph in
the physical region, and how results for a whole set of 
different numerical values for the  invariants can be produced and plotted in an automated way.
We will use the two-loop diagram shown in Fig.~\ref{fig:P126} as an example. 
Numerical results for this diagram have been produced in \cite{Ferroglia:2003yj,Bonciani:2003hc}, 
and  an analytical result can be found in \cite{Davydychev:2003mv}, 
where the diagram is called $P_{126}$. 

\begin{figure}[ht]
\begin{center}
\begin{minipage}{19pc}
\includegraphics[width=25pc]{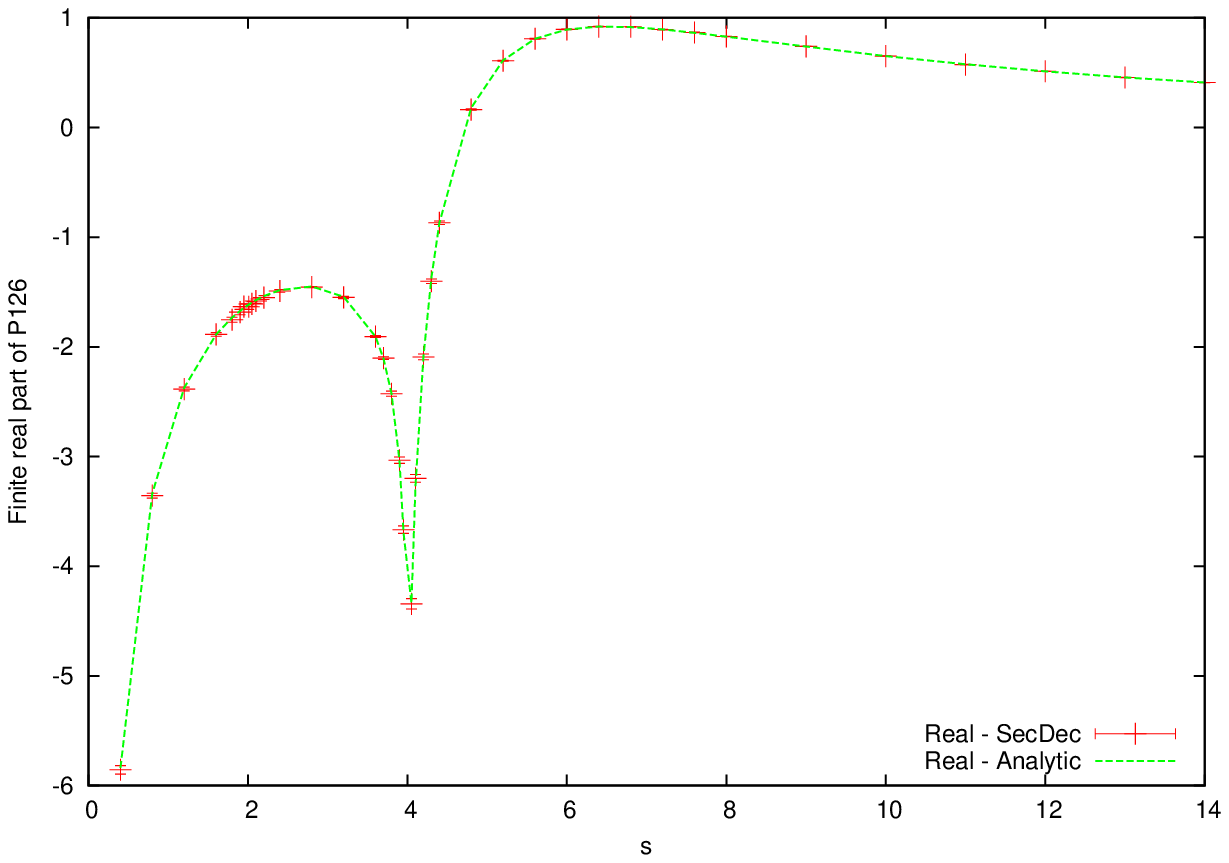}
\end{minipage}\hspace{2pc}%
\begin{minipage}{19pc}
\includegraphics[width=25pc]{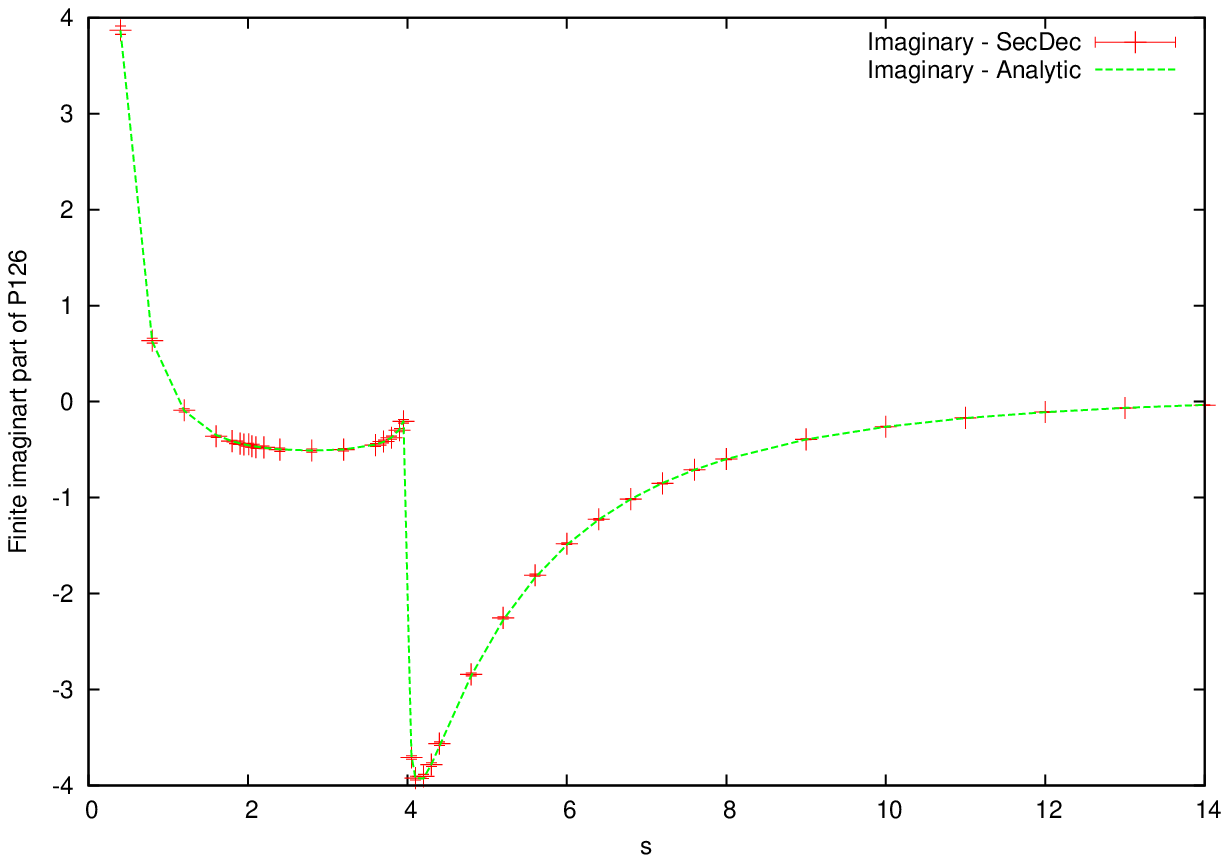}
\end{minipage} 
\caption{\label{fig:P126result}
Comparison of analytic and numerical results for the diagram $P_{126}$ using $m^2=1$.}
\end{center}
\end{figure}

The template  file {\tt templateP126.m} in the {\tt demos} subdirectory 
contains the following lines:\\
{\tt proplist}={\small\{\{ms[1],\{3,4\}\},\{ms[1],\{4,5\}\},\{ms[1],\{5,3\}\},\{0,\{1,2\}\},\{0,\{1,4\}\},\{0,\{2,5\}\}\}};\\
{\tt onshell}=\{ssp[1]$\to$ 0,ssp[2]$\to$ 0,ssp[3]$\to$ sp[1,2]\};\\
where each entry in {\tt proplist} corresponds to a propagator of the diagram; the first entry is the mass of the
propagator, and the second entry contains the labels of the two vertices which the propagator connects. 
The labels for the vertices are as shown in Fig.\,\ref{fig:P126}.
Note that if an external momentum $p_k$ is flowing into the vertex, the vertex must also have the label $k$.
For vertices containing only internal propagators the labeling is arbitrary.
The on-shell conditions in the above example state that
$p_1^2=p_2^2=0,\,p_3^2=s_{12}=s$.
Results for the $\eps^0$ part of graph $P_{126}$ are shown in Fig.\,\ref{fig:P126result}. 

To run this example, from the {\tt loop} directory, issue the command
{\tt ./launch -d demos -p paramP126.input -t templateP126.m}.
The timings for the finite part and a relative accuracy of about 1\%, 
using \textsc{Cuba}-3.0\,\cite{Agrawal:2011tm}, are around 100 secs
for a typical point far from the $s=4m^2$ threshold on an Intel(R) Core(TM) i7 CPU at 2.67GHz with eight cores. 
For a point close to threshold ($s/m^2=3.9$), the timings are similar.

\subsection*{Producing data files for sets of numerical values}
To loop over a set of numerical values for the invariants  $s$ and $m^2$
once the C++ files are created, issue the command\\
{\tt perl multinumerics.pl -d demos -p multiparamP126.input}. 
This will run the numerical integrations for the values of $s$ and $m^2$ specified
in the file {\tt demos/multiparamP126.input}. 
The files containing the results will be found in {\tt demos/2loop/P126}, and the
files {\tt p-2.gpdat, p-1.gpdat} and {\tt p0.gpdat} will contain the  
data files for each point, corresponding to the coefficients of 
$\eps^{-2},\eps^{-1}$ and $\eps^0$ respectively. 
These files can be used to plot the results against the analytic results using gnuplot. 
This will produce the files {\tt P126R0.ps, P126I0.ps} which will look like Fig.~\ref{fig:P126result}.

\clearpage

\subsubsection{Non-planar massive two-loop four-point functions}
\label{sec:ggtt}

\subsubsection*{The graph $B_6^{NP}$}
\begin{figure}[ht]
\begin{center}
\unitlength=1mm
\begin{picture}(100,50)
\put(10,5){\includegraphics[width=5.5cm]{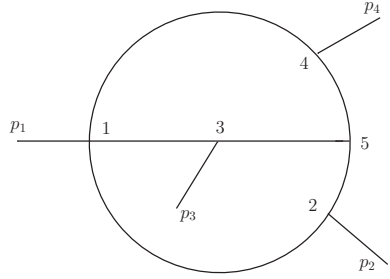}}
\end{picture}
\end{center}
\caption{The non-planar 6-propagator graph $B_6^{NP}$.}
\label{fig:Bnp6}
\end{figure}
\begin{figure}[htb]
\begin{center}
\begin{minipage}{17pc}
\includegraphics[width=15pc,angle=-90]{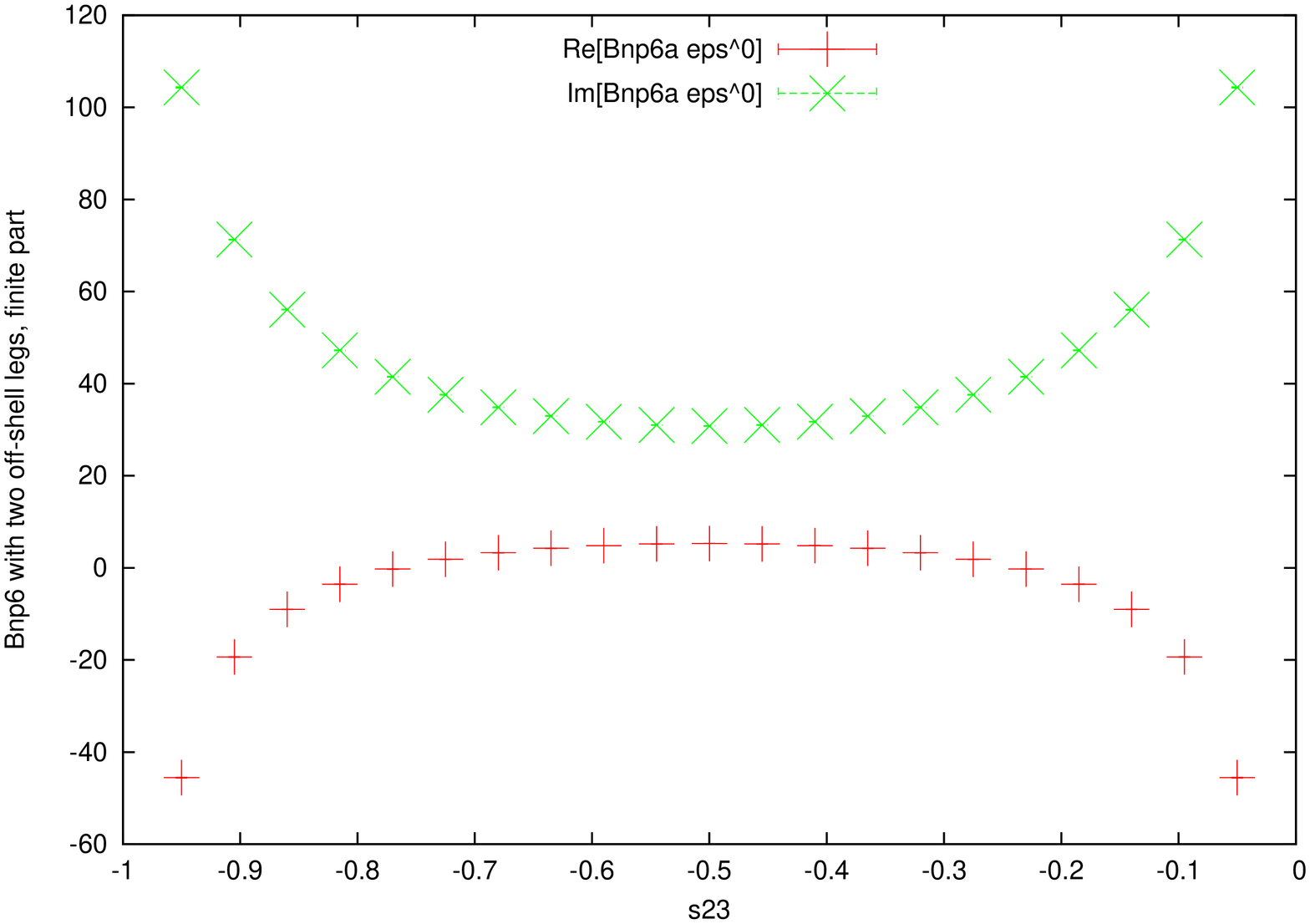}
\end{minipage}\hspace{2pc}%
\begin{minipage}{17pc}
\includegraphics[width=15pc,angle=-90]{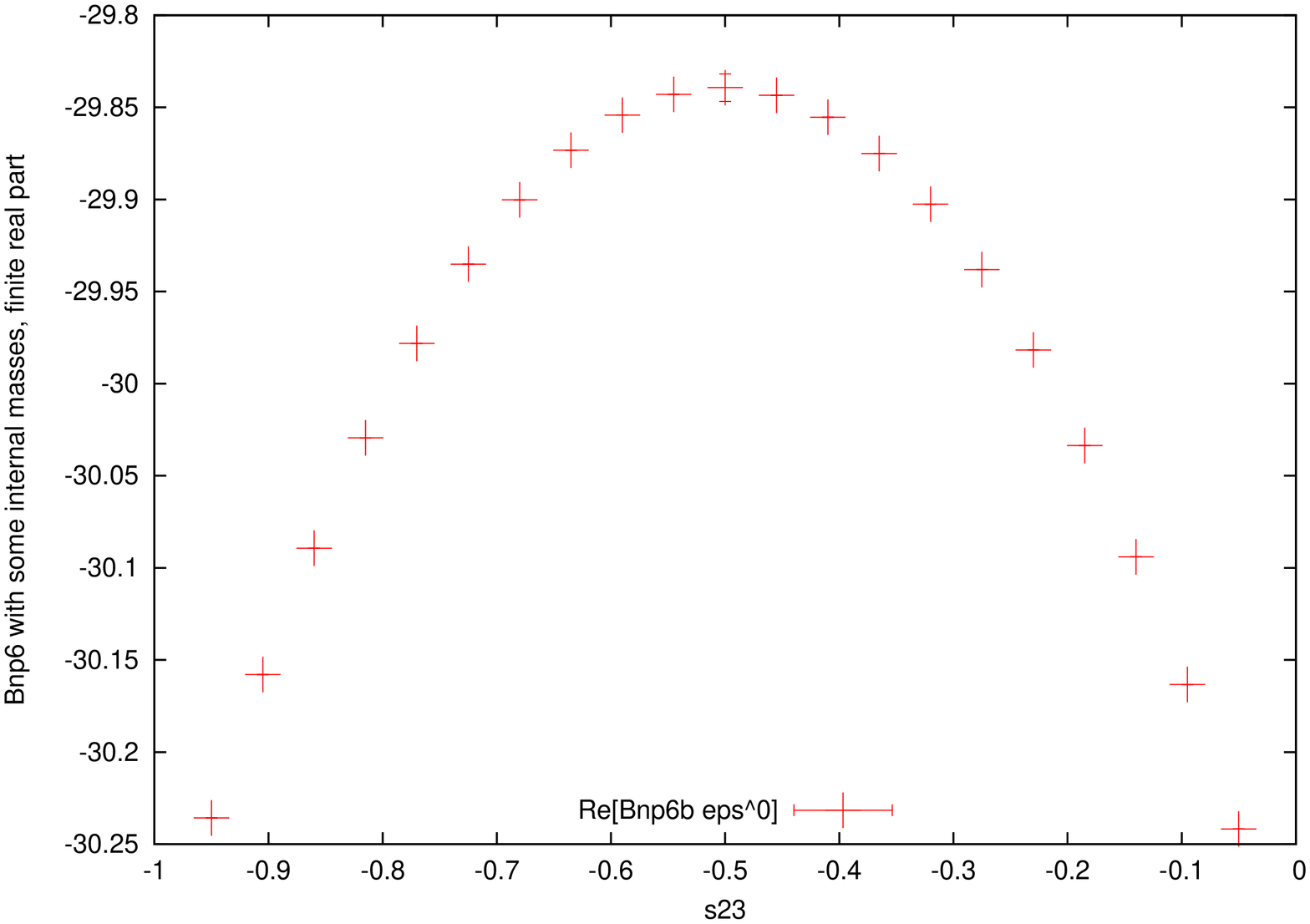}
\end{minipage} 
\caption{\label{fig:Bnp6result}
Results for the finite part of the graph $B_6^{NP}$ (a) with $p_1^2$ and $p_2^2$ off-shell, (b) 
with $m_1=m_2=m_5=m_6=0.25,m_3=m_4=0.$ The imaginary part of  $B_6^{NP,b}$ is zero in the range shown above.}
\end{center}
\end{figure}
\begin{figure}[htb]
\begin{center}
\unitlength=1mm
\includegraphics[width=7.cm,angle=-90]{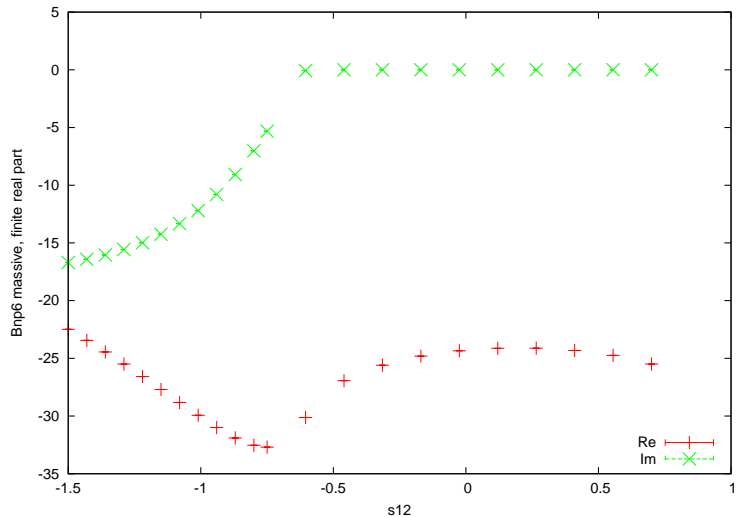}
\end{center}
\caption{The non-planar 6-propagator graph $B_6^{NP,b}$ as a function of $s_{12}$ 
in a region containing a threshold.}
\label{fig:Bnp6threshold}
\end{figure}

Next, we consider the non-planar 6-propagator two-loop four-point diagram shown in Fig.\,\ref{fig:Bnp6}.
For light-like legs and massless propagators, the analytic result has been calculated in \cite{Tausk:1999vh}, 
where the graph is called $B_6^{NP}$.
Here we give results for this topology for the cases where\\
 (a) $p_1^2$ and $p_2^2$ are off-shell\\
 (b) $m_1=m_2=m_5=m_6\not=0,m_3=m_4=0.$\\
It is interesting to note that  $B_6^{NP,a}$ with $p_1^2$ and $p_2^2$ being off-shell contains poles 
starting from $1/\eps^4$, while for light-like legs the leading pole is only $1/\eps^2$, due to 
cancellations related to the high symmetry of the graph.
For $B_6^{NP,b}$, i.e. the graph with $m_1=m_2=m_5=m_6\not=0$, the leading pole is $1/\eps$. 
Results for the finite parts of $B_6^{NP,a}$ and $B_6^{NP,b}$ are shown in 
Figs.\,\ref{fig:Bnp6result} and \ref{fig:Bnp6threshold}.
As in \cite{Tausk:1999vh}, an overall prefactor of 
$\Gamma(1+2\eps)\Gamma(1-\eps)^3/\Gamma(1-3\eps)/(1+4\eps)$ has been extracted.
For Fig.\,\ref{fig:Bnp6result} we have used the numerical values $s_{12}=3$ 
while scanning over $s_{23}$. For all the values given, $s_{13}$ is determined by the physical constraint 
$s_{12}+s_{13}+s_{23}=p_1^2+p_2^2$.
For $B_6^{NP,a}$  we have set $p_1^2=p_2^2=1$, while for $B_6^{NP,b}$, $m_1=m_2=m_5=m_6=0.25$ has been used.
Fig.\,\ref{fig:Bnp6threshold} shows $B_6^{NP,b}$ as a function of $s_{12}$
with $s_{23}=-0.4$ and  $m_1=m_2=m_5=m_6=0.25$. 
The numerical accuracy is about one permil, therefore the error bars are barely seen in the figures.


\subsubsection*{The graph $J^{NP}$}

\begin{figure}[ht]
\begin{center}
\includegraphics[width=7.cm]{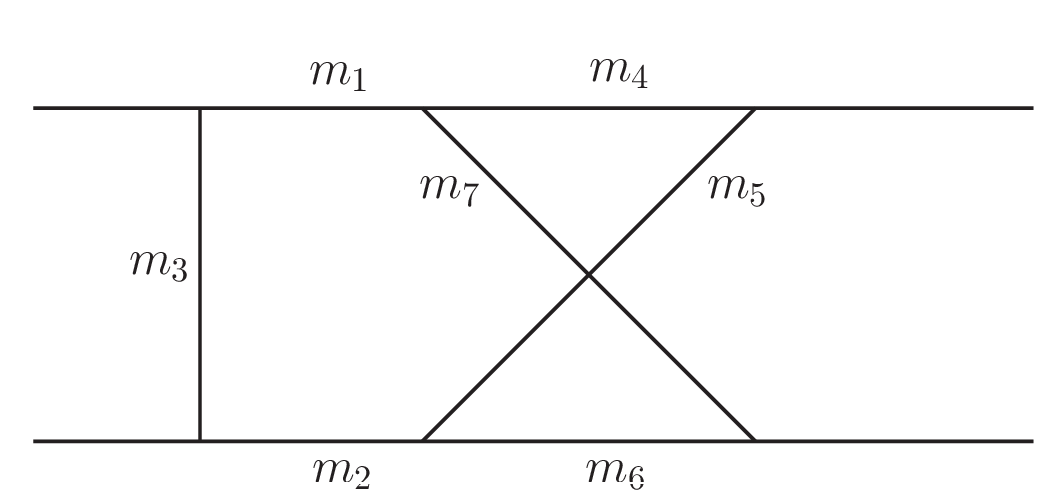}
\end{center}
\caption{Labeling of the masses for the non-planar graph $J^{NP}$.}
\label{fig:JapNP}
\end{figure}

In this example we consider a 7-propagator non-planar two-loop box integral where all propagators are massive, 
using $m_1=m_2=m_5=m_6=m$,  $m_3=m_4=m_7=M$, $p_1^2=p_2^2=p_3^2=p_4^2=m^2$. 
The labelling is as shown in Fig.\,\ref{fig:JapNP}. 

\begin{figure}[ht]
\begin{center}
\includegraphics[width=9.5cm, angle=0]{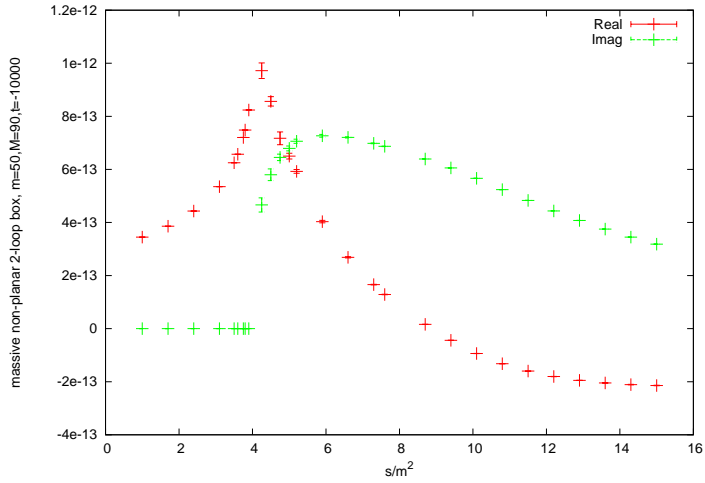}
\end{center}
\caption{Results for a non-planar 7-propagator graph $J^{NP}$ with all propagators massive using $m=50$, $M=90$ and $t=-10^4$.} 
\label{fig:JapNPresults}
\end{figure}

Numerical results for this integral have been calculated in \cite{Yuasa:2011ff} using a
method based on extrapolation in the $i\delta$ parameter.
Our results for $m=50, M=90, s_{23}=-10^4$ are shown in Fig.\,\ref{fig:JapNPresults} and agreement with 
ref. \cite{Yuasa:2011ff} has been verified.
The timings for the longest subfunction (both real and imaginary part) 
with a relative accuracy of one permil vary between about 20 seconds
for a point far from threshold and about 500 seconds close to threshhold.

\clearpage
\subsubsection*{The graph $ggtt_1$}

\begin{figure}[ht]
\begin{center}
\includegraphics[width=6.cm]{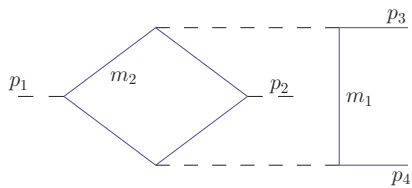}
\end{center}
\caption{Non-planar graph ocurring in the calculation of $gg\to t\bar{t}$ at NNLO.
Blue (solid) lines denote massive particles.}
\label{fig:ggtt}
\end{figure}

The graph shown in Fig.\,\ref{fig:ggtt} occurs in the calculation 
of the two-loop corrections to heavy quark production. 
Numerical results for the two-loop amplitude in the $q\bar{q}$ initiated channel have been calculated 
in \cite{Czakon:2008zk}. 
Analytic results in the $q\bar{q}$ channel and for some colour structures in the $gg$ channel
have been calculated in \cite{Bonciani:2008az,Bonciani:2009nb,Bonciani:2010mn}.
Numerical results for the amplitude in the $gg$ channel
in the approximation $s\gg m_t^2$ have been calculated in \cite{Czakon:2007ej,Czakon:2007wk}.

\begin{figure}[ht]
\begin{center}
\unitlength=1mm
\includegraphics[width=8.5cm,angle=0]{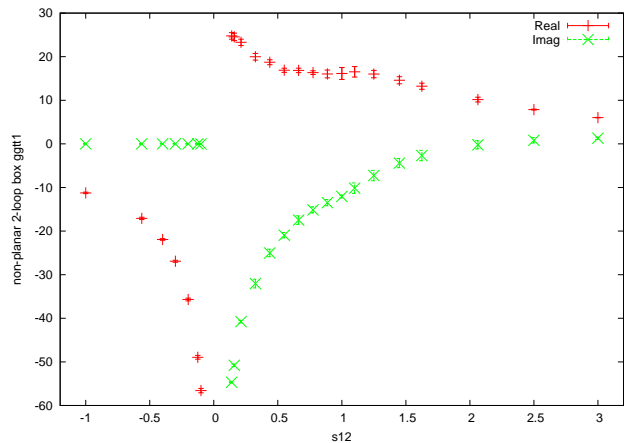}
\end{center}
\caption{Results for the non-planar graph $ggtt_1$.}
\label{fig:ggtt1results}
\end{figure}

For the individual graph shown in Fig.\,\ref{fig:ggtt}, 
numerical results at Euclidean points have been given in \cite{Carter:2010hi}. 
Here we give numerical results for the non-planar master integral $ggtt_1$ 
in the non-Euclidean region. 
For the results shown in Fig.~\ref{fig:ggtt1results}, we used 
$s_{23}=-0.4,p_3^2=p_4^2=0.25,m_1^2=m_2^2=0.25,s_{13}=-s_{12}-s_{23}+p_3^2+p_4^2$.
The analytical result for this master integral is not known yet.

\subsection{Defining implicit functions}

\label{example:dummy}
This example demonstrates the ability to leave certain functions implicit until numerical
integration. Suppose we want to integrate
\begin{equation}
f(\vec{x})=(x_1+x_2)^{-2-2\epsilon}x_3^{-1-4\epsilon}dum_1(x_1,x_2,x_3,x_4)^{1+\eps}dum_2(x_2,x_4)^{2-6\epsilon}cut(x_3)
\nonumber
\end{equation}
with
\begin{eqnarray*}
\label{eq:dummy2} 
dum_1(x_1,x_2,x_3,x_4)&=&2+x_1^2+x_2^3+x_3^4+x_4^5+4x_1x_2x_3x_4-x_1^2x_2^3x_3^4x_4^5\\
dum_2(x_2,x_4)&=&x_2^2+x_4^2+\beta^2+4x_2x_4-\sqrt{x_2x_4\beta}+3x_2^2 x_4^2\;,
\end{eqnarray*}
where $\beta$ is a symbol defined in the parameter file. $dum_1$ and $dum_2$ 
should always be functions which cannot increase the singular behaviour of the
integrand, and so quantitative knowledge of their exact form is not required to guide the
decomposition. Thus they can be left implicit, and only introduced at the numerical integration
stage. The function $cut(x)$ is already defined by default as $cut(x)\equiv\Theta(x-cutval)$, 
where the value of $cutval$ is given in the parameter file.
The command {\it ../launch -p
paramdummy.input -t templatedummy.m} from the folder {\tt general/demos} runs this example.
The fortran files containing the explicit form of the functions $dum_1$, $dum_2$, $cut$ are found
in {\tt demos/testdummy}.
Another simple example for a dummy function 
would be a jet function as used in the {\small JADE} algorithm in $e^+e^-$
annihilation, as described e.g. in  \cite{Bethke:1988zc}.
Input files for such an example are 
{\tt params23s35JADE.input, templates23s35JADE.m} in the {\tt general/demos}
directory.

%% file: appendix.tex
We list here all possible input parameters for the parameter file {\tt*.input} and the Mathematica input file {\tt*.m}. 
These two files serve to define the integrand and the parameters for the numerical integration.
We describe here the example of loop diagrams; the input files in the subdirectory {\tt general}
to compute more general parametric functions is very similar.
\subsection{Program input parameters}
This input file should be called {\tt *.input}. The following parameters can be specified
\begin{description}
\item[subdir]
{\it subdir} specifies the name of the subdirectory to which the graph should be written to. If not yet existent, it will be created. The specified {\it subdir} contains the directory specified in outputdir.
\item[outputdir]
The name for the desired output directory can be given here. 
If {\it outputdir} is not specified, the default directory for the output will have the graph name 
(see below) appended to the directory {\it subdir}, otherwise specify the full path for the 
Mathematica output files here.\\
The output directory will contain all the files produced during the decomposition, subtraction, 
expansion and numerical integration, and the results. 
The output of the decomposition into sectors is found in the outputdir directly. 
The functions from subtraction and expansion and the respective files for numerical integration are 
found in subdirectories. The latter are named with the pole structure and contain subdirectories named 
with the respective Laurent coefficient. 
\item[graph]
The name of the diagram or parametric function to be computed is specified here. 
The graph name can contain underscores and numbers, but should not contain commas.
\item[propagators]
The number of propagators the diagram has is specified here (mandatory).
\item[legs]
The number of external legs the diagram has is specified here (mandatory).
\item[loops]
The number of loops the diagram has is specified here (mandatory).
\item[cutconstruct]
If the graph to be computed corresponds to a scalar integral, 
the integrand (F and U) can be constructed via topological cuts.
In this case set cutconstruct=1, the default is =0. 
If cutconstruct is switched on, the input for the graph structure (*.m file) 
is just a list of labels connecting vertices, 
as explained in Section \ref{subsec:cutconstruct} and \ref{subsec:graphm}.
\item[epsord]
The order to which the Laurent series in $\epsilon$ should be expanded, starting from $\epsilon^{-maxpole}$, can be specified here. 
The default is epsord=0 where the Laurent series is cut after finite part $\epsilon^0$. 
If epsord is set to a negative value, only the pole coefficients up to this order will be computed.
\item[prefactorflag]
Possible values for the {\it prefactorflag} are 0 (default), 1 and 2. 
\begin{itemize}
\item 0: The default prefactor $(-1)^{N}\,\Gamma[N-Nloops*Dim/2]$ is factored out of the numerical result.
\item 1: The default prefactor $(-1)^{N}\,\Gamma[N-Nloops*Dim/2]$ is included in the numerical result.
\item 2: Give the desired prefactor to be factored out in {\it prefactor}. 
\end{itemize}
\item[prefactor]
If option 2 has been chosen in the {\it prefactorflag}, write down the desired {\it prefactor} in Mathematica syntax here. In combination with options 0 or 1 in the {\it prefactorflag} this entry will be ignored
Use Nn, Nloops and Dim to denote the number of propagators, loops and dimension (4-2eps by default).
\item[IBPflag]
Set {\it IBPflag=0} if integration by parts should not be used, =1 if it should be used. 
{\it IBPflag=2} is designed to use IBP when it is more efficient to do so, and not otherwise. 
Using the integrations by parts method takes more time in the subtraction and expansion step and generally 
results in more functions for numerical integration. 
However, it can be useful if (spurious) poles of type $x^{-2-b\eps}$ are found in the decomposition, 
as it reduces the power of $x$ in the denominator.
\item[compiler]
Set a Fortran compiler (tested with gfortran, ifort, g77) if {\it language=Fortran}. 
Left blank, the default is gfortran.
\item[exeflag]
The {\it exeflag} is set to decide at which stage the program terminates: 
\begin{itemize}
\item 0: The iterated sector decomposition is done and the scripts to do the subtraction, the expansion in epsilon, the creation of the Fortran/C++ files and to launch the numerical integration are created (scripts batch* in the subdirectory {\it graph}) but not run. This can be useful if a cluster is available to run each pole structure on a different node.
\item 1: In addition to the steps done in 0, the subtraction and epsilon expansion is performed and the resulting functions are written to Fortran/C++ files. 
\item 2: In addition to the steps done in 1, all the files needed for the numerical integration are created.
\item 3: In addition to the steps done in 2, the compilation of the Fortran/C++ files is launched to make the executables. 
\item 4: In addition to the steps done in 3, the executables are run, either by batch submission or locally.
\end{itemize}
\item[clusterflag]
The {\it clusterflag} determines how jobs are submitted. Setting {\it clusterflag=0} (default) the jobs will run on
a single machine, setting it =1 the jobs will run on a cluster (a batch system to submit jobs).
\item[batchsystem]
If a cluster is used ({\it clusterflag=1}), this flag should be set to 0 to use the setup 
for the PBS (Portable batch system). If the flag is set to 1
a user-defined setup is activated. Currently this is the submission via {\tt condor}, but the user can 
adapt this to his needs by editing {\tt perlsrc/makejob.pm}. 
\item[maxjobs]
When using a cluster, specify the maximum number of jobs allowed in the queue here.
\item[maxcput]
Specify here the estimated maximal CPU time (in hours). 
This option is used to send a job to a particular queue on a batch system, otherwise it is not important. 
\item[pointname]
The name of the point to calculate is specified here. 
It should be either blank or a string and is useful to label the result files in case of different 
runs for different numerical values of the Mandelstam variables, masses etc.
\item[sij]
The values for Mandelstam invariants $s_{ij}=(p_i+p_j)^2$ in numbers are specified here (mandatory). 
The $s_{ij}$ should be $\leq 0$ in the Euclidean region. 
\item[pi2]
The off-shell legs $p_1^2$, $p_2^2$,... are specified here (mandatory). $p_i^2$ should be $\leq 0$ in the Euclidean region.
\item[ms2]
Specify the masses of propagators $m_1^2$, $m_2^2$,... 
here using the notation ms[i] for  $m_i^2$ (mandatory). The masses should not be complex numbers. 
\item[integrator]
The program for numerical integration can be chosen here. BASES ({\it integrator=0}) can only be used in the Fortran 
version. Vegas ({\it integrator=1}), Suave ({\it integrator=2}), Divonne ({\it integrator=3}, default) and Cuhre ({\it integrator=4}) 
are part of the \textsc{Cuba} library and can be used in both the Fortran and the C++ version. 
In practice, Divonne usually gives the fastest results when using the C++ version. 
In the following we therefore concentrate on the adjustment of the parameters needed for numerical 
integration using Divonne. For more details about the \textsc{Cuba} parameters we refer to \cite{Hahn:2004fe}.
\item[cubapath]
The path to the \textsc{Cuba} library can be specified here. The default directory is {\it [your path to SecDec]/Cuba-3.0}. 
\textsc{Cuba}-3.0 is the newest version of the \textsc{Cuba} library and uses parallel processing during the numerical 
evaluation of the integral. The older version (\textsc{Cuba}-2.1) is still supported and can be used.
\item[maxeval]
Separated by commas and starting with the lowest order coefficient in $\epsilon$, 
specify the maximal number of evaluations to be used by the numerical integrator for each order in $\epsilon$. 
If {\it maxeval} is not equal to {\it mineval}, the maximal number of evaluations does not have to be reached.
\item[mineval]
Separated by commas and starting with the lowest order coefficient in $\epsilon$, 
specify the number of evaluations which should at least be done before the numerical integrator returns a result. 
The default is 0. 
\item[epsrel]
Separated by commas and starting with the lowest order coefficient in $\epsilon$, 
specify the desired relative accuracy for the numerical evaluation.
\item[epsabs]
Separated by commas and starting with the lowest order coefficient in $\epsilon$, 
specify the desired absolute accuracy for the numerical evaluation. 
This becomes useful in the cases where the integrated result is close to zero.
\item[cubaflags]
Set the cuba verbosity flags. The default is 2 which means, the \textsc{Cuba} input parameters and other useful 
information, e.g. about numerical convergence, are echoed during numerical integration.
\item[key1]
Separated by commas and starting with the lowest order coefficient in $\epsilon$, 
specify {\it key1} which determines the sampling to be used for the partitioning phase in Divonne. 
With a positive {\it key1}, a Korobov quasi-random sample of {\it key1} points is used.
A {\it key1} of about 1000 (default) usually is a good choice. 
\item[key2]
Separated by commas and starting with the lowest order coefficient in $\epsilon$, 
specify {\it key2} which determines the sampling to be used for the final integration phase in Divonne. 
With a positive {\it key2}, a Korobov quasi-random sample is used. 
The default is {\it key2=1} which means, the number of points needed to reach the prescribed accuracy is estimated 
by Divonne. 
\item[key3]
Separated by commas and starting with the lowest order coefficient in $\epsilon$, 
specify the {\it key3} to be used for the refinement phase in Divonne. 
Setting {\it key3=1} (default), each subregion is split once more.
\item[maxpass]
Separated by commas and starting with the lowest order coefficient in $\epsilon$, 
specify how good the convergence has to be during the partitioning phase until the program passes 
on to the main integration phase. 
A {\it maxpass} of 3 (default) is usually sufficient to get a quick and good result.  
\item[border]
Separated by commas and starting with the lowest order coefficient in $\epsilon$, 
specify the border for the numerical integration. 
The points in the interval $[0,border]$ and $[1-border,1]$ are not included in the integration 
but are extrapolated from points further from the endpoints. This can be useful if the integrand 
is known to be peaked at endpoints of the integration variables.
\item[maxchisq]
Separated by commas and starting with the lowest order coefficient in $\epsilon$, 
specify the maximally allowed $\chi^2$ at the end of the numerical integration. 
\item[mindeviation]
Separated by commas and starting with the lowest order coefficient in $\epsilon$, 
specify the deviation two sample averages in one region can show without being treated any further.\vspace{10pt}\\

These parameters are advanced options
\item[primarysectors]
Specify a list of primary sectors to be treated here. If left blank, {\it primarysectors} defaults to all, i.e. 1 to 
the number of propagators, will be taken. This option is useful if a diagram has symmetries such that some primary 
sectors yield the same result. 
\item[multiplicities]
Specify the {\it multiplicities} of the primary sectors listed above. List the {\it multiplicities} in same order as 
the corresponding sectors above. If left blank, default multiplicities (=one) are set automatically.
\item[infinitesectors]
A list of primary sectors to be redone differently because they lead to infinite recursion can be specified here. 
{\it infinitesectors} must be left empty for the default strategy to be applied.
\item[togetherflag]
This flag defines whether to integrate subsets of functions for each pole order separately {\it togetherflag=0}(default) 
or to sum all functions for a certain pole order and then integrate {\it togetherflag=1}. The latter will allow 
cancellations between different functions and thus give a more realistic error, but should not be used for complicated 
diagrams where the individual functions are large already.
\item[editor]
Choose here which editor should be used to display the result. 
If editor=none is set, the full result will not be displayed in an editor window at the end of the calculation.
\item[grouping]
If the {\it togetherflag} is set to 0, it could still be useful to first sum a few functions before integration.
The number of bytes you set with grouping=\#bytes decides how many functions
f*.f or f*.cc are first summed and 
only then integrated with the numerical integrator. 
If you set grouping=0 all functions f*.f resp. f*.cc are integrated separately. 
In practice, a grouping=0 has proven to lead to faster convergence and more accurate results. 
However, if you consider integrals which show large cancellations within the different functions f*.cc, 
it might be useful to use a grouping$\neq0$. The log files *results*.log in the results directory contain 
the results from the individual integration, where the user can see if there are large cancellations between 
the individual functions.
\item[language]
For one-scale diagrams or diagrams with purely Euclidean kinematics language=fortran or language=Cpp (default) 
can be chosen, where the Cpp stands for C++. \\
In all other cases, especially when using contourdef=True, language=Cpp is used, as the deformation of the contour 
which is needed for these problems is only implemented in C++. 
\item[rescale]
If all invariants are very small or very large it is useful to rescale them to reach faster convergence during 
numerical integration. The rescaling (scaling out the largest invariant
in the numerical integration part) can be switched on with {\it rescale=1} and switched off when set to 0. 
If switched on, it is not  possible to set explicit values of any non-zero invariants in the Mathematica input file {\tt template*.m}.
\item[contourdef]
For multi-scale problems resp. diagrams with non-Euclidean kinematics, set contourdef=True (default is False). 
In this case, a deformation of the integration contour in the form of Eq.\,(\ref{eq:condef}) is done. 
In addition to the functions f*.cc to be integrated, auxiliary files (g*.cc) are written which serve to
optimize the deformation for each integrand function.
\item[lambda]
Here, you can set the initial lambda $\lambda$ for the deformation of Eq.\,(\ref{eq:condef}). 
Without any knowledge about the characteristics of the integrand, {\it lambda=1.0} should be a good choice. 
If the diagram contains mostly massless propagators and  light-like legs, 
it can be useful to choose the initial $\lambda$  larger (e.g.{\it lambda=5.0}), in order to 
compensate for cases where the remainders of the IR subtraction lead to large cancellations 
for $x_i\to 0$. 
For diagrams with mostly massive propagators the initial lambda can be chosen smaller (e.g.{\it lambda=0.1}).    
\item[smalldefs]
If the integrand is expected to be oscillatory and hence sensitive to small changes in the 
deformation parameter $\lambda$, {\it smalldefs} should be set to 1 (default is 0). 
If switched on, the argument of each subsector function ${\cal F}$ is minimized.
\item[largedefs]
If the integrand is expected to  have (integrable) endpoint singularities 
at $x_j=0$ or 1, the deformation should be large in order to move the contour away 
from the problematic region. If {\it largedefs=1}, the 
program tries to enlarge the deformation at the endpoints.
The default is {\it largedefs=0}. 
\end{description}

\subsection{Input for the definition of the integrand}
\label{subsec:graphm}
This Mathematica input file should be called {\tt *.m}. The following parameters can be specified
\begin{description}
\item[momlist]
If {\it cutconstruct=0} is set in the input file, specify the names of the loop momenta here.
\item[proplist]
Specify the diagram topology here (mandatory). The syntax for {\it cutconstruct=1} is described in 
Section \ref{subsec:cutconstruct}. 
If {\it cutconstruct=0} has been chosen, the propagators have to be given explicitly.
An example propagator list could be {\it proplist=\{k\^\,2-ms[1],(k+p1)\^\,2-ms[1]\}} with the loop momentum $k$, 
the propagator mass $m_1^2$ and external momentum $p_1$. 
\item[numerator]
If present, specify the numerator of the integrand here. 
If not given, a {\it numerator=\{1\}} is assumed. Please note that the option 
{\it cutconstruct=1} is not available in combination with numerator functions. 
\item[powerlist]
As an option, the propagator powers (e.g. if different from one) can be set here.
\item[onshell]
Specify invariant replacements here. The kinematic invariants can be assigned values (e.g. ssp[1]$\to$0) 
or relations between the invariants can be set (e.g. ssp[1]$\to$sp[1,3]). 
This option can not be used in combination with {\it rescale=1}.
\item[Dim]
Set the space-time dimension. The default is {\it Dim=4-2\,eps} and the symbol 
for the regulator $\epsilon$ must remain the same.
\end{description}

%% file: condef-main.bbl
\begin{thebibliography}{10}

\bibitem{Binoth:2000ps}
T.~Binoth and G.~Heinrich.
\newblock An automatized algorithm to compute infrared divergent multi-loop
  integrals.
\newblock {\em Nucl. Phys.}, B585:741--759, 2000.

\bibitem{Roth:1996pd}
M.~Roth and Ansgar Denner.
\newblock High-energy approximation of one-loop {F}eynman integrals.
\newblock {\em Nucl. Phys.}, B479:495--514, 1996.

\bibitem{Hepp:1966eg}
Klaus Hepp.
\newblock Proof of the {B}ogolyubov-{P}arasiuk theorem on renormalization.
\newblock {\em Commun. Math. Phys.}, 2:301--326, 1966.

\bibitem{Carter:2010hi}
Jonathon Carter and Gudrun Heinrich.
\newblock {SecDec: A general program for sector decomposition}.
\newblock {\em Comput.Phys.Commun.}, 182:1566--1581, 2011.

\bibitem{Bogner:2007cr}
Christian Bogner and Stefan Weinzierl.
\newblock {Resolution of singularities for multi-loop integrals}.
\newblock {\em Comput. Phys. Commun.}, 178:596--610, 2008.

\bibitem{Smirnov:2008py}
A.V. Smirnov and M.N. Tentyukov.
\newblock {Feynman Integral Evaluation by a Sector decomposiTion Approach
  (FIESTA)}.
\newblock {\em Comput.Phys.Commun.}, 180:735--746, 2009.

\bibitem{Smirnov:2009pb}
A.V. Smirnov, V.A. Smirnov, and M.~Tentyukov.
\newblock {FIESTA 2: Parallelizeable multiloop numerical calculations}.
\newblock {\em Comput.Phys.Commun.}, 182:790--803, 2011.

\bibitem{Gluza:2010rn}
Janusz Gluza, Krzysztof Kajda, Tord Riemann, and Valery Yundin.
\newblock {Numerical Evaluation of Tensor Feynman Integrals in Euclidean
  Kinematics}.
\newblock {\em Eur.Phys.J.}, C71:1516, 2011.

\bibitem{Ueda:2009xx}
Takahiro Ueda and Junpei Fujimoto.
\newblock {New implementation of the sector decomposition on FORM}.
\newblock {\em PoS}, ACAT08:120, 2008.

\bibitem{Heinrich:2008si}
Gudrun Heinrich.
\newblock {Sector Decomposition}.
\newblock {\em Int. J. Mod. Phys.}, A23:1457--1486, 2008.

\bibitem{Anastasiou:2010pw}
Charalampos Anastasiou, Franz Herzog, and Achilleas Lazopoulos.
\newblock {On the factorization of overlapping singularities at NNLO}.
\newblock {\em JHEP}, 1103:038, 2011.
\newblock 36 pages.

\bibitem{Pilipp:2008ef}
Volker Pilipp.
\newblock {Semi-numerical power expansion of Feynman integrals}.
\newblock {\em JHEP}, 0809:135, 2008.

\bibitem{Czakon:2004wm}
M.~Czakon, J.~Gluza, and T.~Riemann.
\newblock Master integrals for massive two-loop {B}habha scattering in {QED}.
\newblock {\em Phys. Rev.}, D71:073009, 2005.

\bibitem{Denner:2004iz}
Ansgar Denner and S.~Pozzorini.
\newblock An algorithm for the high-energy expansion of multi-loop diagrams to
  next-to-leading logarithmic accuracy.
\newblock {\em Nucl. Phys.}, B717:48--85, 2005.

\bibitem{Denner:2008yn}
Ansgar Denner, Bernd Jantzen, and Stefano Pozzorini.
\newblock {Two-loop electroweak next-to-leading logarithms for processes
  involving heavy quarks}.
\newblock {\em JHEP}, 0811:062, 2008.

\bibitem{Czakon:2010td}
M.~Czakon.
\newblock {A novel subtraction scheme for double-real radiation at NNLO}.
\newblock {\em Phys.Lett.}, B693:259--268, 2010.
\newblock 14 pages, 3 figures, matches published version, includes new name for
  the scheme, extended discussion of massless final states, and some new
  references.

\bibitem{Czakon:2011ve}
M.~Czakon.
\newblock {Double-real radiation in hadronic top quark pair production as a
  proof of a certain concept}.
\newblock {\em Nucl.Phys.}, B849:250--295, 2011.
\newblock 44 pages, 10 figures.

\bibitem{Boughezal:2011jf}
Radja Boughezal, Kirill Melnikov, and Frank Petriello.
\newblock {A subtraction scheme for NNLO computations}.
\newblock {\em Phys.Rev.}, D85:034025, 2012.
\newblock 13 pages.

\bibitem{Bolzoni:2010bt}
Paolo Bolzoni, Gabor Somogyi, and Zoltan Trocsanyi.
\newblock {A subtraction scheme for computing QCD jet cross sections at NNLO:
  integrating the iterated singly-unresolved subtraction terms}.
\newblock {\em JHEP}, 1101:059, 2011.
\newblock 83 pages, one reference added, typos corrected, agrees with published
  version.

\bibitem{Heinrich:2002rc}
Gudrun Heinrich.
\newblock A numerical method for {NNLO} calculations.
\newblock {\em Nucl. Phys. Proc. Suppl.}, 116:368--372, 2003.

\bibitem{Anastasiou:2003gr}
Charalampos Anastasiou, Kirill Melnikov, and Frank Petriello.
\newblock A new method for real radiation at {NNLO}.
\newblock {\em Phys. Rev.}, D69:076010, 2004.

\bibitem{GehrmannDeRidder:2003bm}
A.~Gehrmann-De~Ridder, T.~Gehrmann, and G.~Heinrich.
\newblock Four-particle phase space integrals in massless {QCD}.
\newblock {\em Nucl. Phys.}, B682:265--288, 2004.

\bibitem{Binoth:2004jv}
T.~Binoth and G.~Heinrich.
\newblock Numerical evaluation of phase space integrals by sector
  decomposition.
\newblock {\em Nucl. Phys.}, B693:134--148, 2004.

\bibitem{Anastasiou:2004qd}
Charalampos Anastasiou, Kirill Melnikov, and Frank Petriello.
\newblock Real radiation at {NNLO}: $e^+ e^- \to 2$ jets through
  {O}$(\alpha_s^2)$.
\newblock {\em Phys. Rev. Lett.}, 93:032002, 2004.

\bibitem{Passarino:2001jd}
Giampiero Passarino and Sandro Uccirati.
\newblock {Algebraic numerical evaluation of Feynman diagrams: Two loop
  selfenergies}.
\newblock {\em Nucl.Phys.}, B629:97--187, 2002.

\bibitem{Ferroglia:2002mz}
Andrea Ferroglia, Massimo Passera, Giampiero Passarino, and Sandro Uccirati.
\newblock {All purpose numerical evaluation of one loop multileg Feynman
  diagrams}.
\newblock {\em Nucl.Phys.}, B650:162--228, 2003.

\bibitem{Ferroglia:2003yj}
Andrea Ferroglia, Massimo Passera, Giampiero Passarino, and Sandro Uccirati.
\newblock {Two loop vertices in quantum field theory: Infrared convergent
  scalar configurations}.
\newblock {\em Nucl.Phys.}, B680:199--270, 2004.

\bibitem{Actis:2004bp}
Stefano Actis, Andrea Ferroglia, Giampiero Passarino, Massimo Passera, and
  Sandro Uccirati.
\newblock {Two-loop tensor integrals in quantum field theory}.
\newblock {\em Nucl.Phys.}, B703:3--126, 2004.

\bibitem{Passarino:2006gv}
Giampiero Passarino and Sandro Uccirati.
\newblock {Two-loop vertices in quantum field theory: Infrared and collinear
  divergent configurations}.
\newblock {\em Nucl.Phys.}, B747:113--189, 2006.
\newblock 62 pages, 15 figures, 16 tables.

\bibitem{Actis:2008ts}
Stefano Actis, Giampiero Passarino, Christian Sturm, and Sandro Uccirati.
\newblock {NNLO Computational Techniques: The Cases H to gamma gamma and H to g
  g}.
\newblock {\em Nucl.Phys.}, B811:182--273, 2009.
\newblock LaTeX, 70 pages, 8 eps figures.

\bibitem{Soper:1999xk}
Davison~E. Soper.
\newblock Techniques for {QCD} calculations by numerical integration.
\newblock {\em Phys. Rev.}, D62:014009, 2000.

\bibitem{Binoth:2002xh}
T.~Binoth, G.~Heinrich, and N.~Kauer.
\newblock A numerical evaluation of the scalar hexagon integral in the physical
  region.
\newblock {\em Nucl. Phys.}, B654:277--300, 2003.

\bibitem{Nagy:2006xy}
Zoltan Nagy and Davison~E. Soper.
\newblock Numerical integration of one-loop {F}eynman diagrams for {N}-photon
  amplitudes.
\newblock {\em Phys. Rev.}, D74:093006, 2006.

\bibitem{Binoth:2005ff}
T.~Binoth, J.~Ph. Guillet, G.~Heinrich, E.~Pilon, and C.~Schubert.
\newblock An algebraic / numerical formalism for one-loop multi-leg amplitudes.
\newblock {\em JHEP}, 10:015, 2005.

\bibitem{Gong:2008ww}
Wei Gong, Zoltan Nagy, and Davison~E. Soper.
\newblock {Direct numerical integration of one-loop Feynman diagrams for
  N-photon amplitudes}.
\newblock {\em Phys. Rev.}, D79:033005, 2009.

\bibitem{Lazopoulos:2007ix}
Achilleas Lazopoulos, Kirill Melnikov, and Frank Petriello.
\newblock {QCD} corrections to tri-boson production.
\newblock {\em Phys. Rev.}, D76:014001, 2007.

\bibitem{Lazopoulos:2008de}
Achilleas Lazopoulos, Thomas McElmurry, Kirill Melnikov, and Frank Petriello.
\newblock {Next-to-leading order QCD corrections to $t \bar{t} Z$ production at
  the LHC}.
\newblock {\em Phys.Lett.}, B666:62--65, 2008.

\bibitem{Becker:2010ng}
Sebastian Becker, Christian Reuschle, and Stefan Weinzierl.
\newblock {Numerical NLO QCD calculations}.
\newblock {\em JHEP}, 1012:013, 2010.

\bibitem{Becker:2011vg}
Sebastian Becker, Daniel Goetz, Christian Reuschle, Christopher Schwan, and
  Stefan Weinzierl.
\newblock {NLO results for five, six and seven jets in electron-positron
  annihilation}.
\newblock {\em Phys.Rev.Lett.}, 108:032005, 2012.
\newblock 5 pages.

\bibitem{Kurihara:2005ja}
Y.~Kurihara and T.~Kaneko.
\newblock {Numerical contour integration for loop integrals}.
\newblock {\em Comput.Phys.Commun.}, 174:530--539, 2006.

\bibitem{Anastasiou:2007qb}
Charalampos Anastasiou, Stefan Beerli, and Alejandro Daleo.
\newblock Evaluating multi-loop {F}eynman diagrams with infrared and threshold
  singularities numerically.
\newblock {\em JHEP}, 05:071, 2007.

\bibitem{Anastasiou:2008rm}
Charalampos Anastasiou, Stefan Beerli, and Alejandro Daleo.
\newblock {The two-loop QCD amplitude gg -> h,H in the Minimal Supersymmetric
  Standard Model}.
\newblock {\em Phys. Rev. Lett.}, 100:241806, 2008.

\bibitem{Beerli:2008zz}
Stefan Beerli.
\newblock {A New method for evaluating two-loop Feynman integrals and its
  application to Higgs production}.
\newblock 2008.
\newblock Ph.D. Thesis (Advisor: Zoltan Kunszt).

\bibitem{deDoncker:2004fb}
E.~de~Doncker, Y.~Shimizu, J.~Fujimoto, and F.~Yuasa.
\newblock {Computation of loop integrals using extrapolation}.
\newblock {\em Comput.Phys.Commun.}, 159:145--156, 2004.

\bibitem{Yuasa:2011ff}
F.~Yuasa, E.~de~Doncker, N.~Hamaguchi, T.~Ishikawa, K.~Kato, et~al.
\newblock {Numerical Computation of Two-loop Box Diagrams with Masses}.
\newblock 2011.

\bibitem{Bauberger:1994by}
S.~Bauberger, Frits~A. Berends, M.~Bohm, and M.~Buza.
\newblock {Analytical and numerical methods for massive two loop selfenergy
  diagrams}.
\newblock {\em Nucl.Phys.}, B434:383--407, 1995.

\bibitem{Fujimoto:1995ev}
Junpei Fujimoto, Yoshimitsu Shimizu, Kiyoshi Kato, and Toshiaki Kaneko.
\newblock {Numerical approach to two loop three point functions with masses}.
\newblock {\em Int.J.Mod.Phys.}, C6:525--530, 1995.

\bibitem{Caffo:2002ch}
Michele Caffo, H.~Czyz, and E.~Remiddi.
\newblock {Numerical evaluation of the general massive 2 loop sunrise selfmass
  master integrals from differential equations}.
\newblock {\em Nucl.Phys.}, B634:309--325, 2002.

\bibitem{Pozzorini:2005ff}
S.~Pozzorini and E.~Remiddi.
\newblock {Precise numerical evaluation of the two loop sunrise graph master
  integrals in the equal mass case}.
\newblock {\em Comput.Phys.Commun.}, 175:381--387, 2006.

\bibitem{Czakon:2005rk}
M.~Czakon.
\newblock {Automatized analytic continuation of Mellin-Barnes integrals}.
\newblock {\em Comput.Phys.Commun.}, 175:559--571, 2006.

\bibitem{Czakon:2008zk}
M.~Czakon.
\newblock {Tops from Light Quarks: Full Mass Dependence at Two-Loops in QCD}.
\newblock {\em Phys.Lett.}, B664:307--314, 2008.

\bibitem{Freitas:2010nx}
Ayres Freitas and Yi-Cheng Huang.
\newblock {On the Numerical Evaluation of Loop Integrals With Mellin-Barnes
  Representations}.
\newblock {\em JHEP}, 1004:074, 2010.

\bibitem{Tarasov:1996br}
O.~V. Tarasov.
\newblock Connection between {F}eynman integrals having different values of the
  space-time dimension.
\newblock {\em Phys. Rev.}, D54:6479--6490, 1996.

\bibitem{Smirnov:2006ry}
Vladimir~A. Smirnov.
\newblock {\em {Feynman integral calculus}}.
\newblock Springer, 2006.

\bibitem{Landau:1959fi}
L.~D. Landau.
\newblock {On analytic properties of vertex parts in quantum field theory}.
\newblock {\em Nucl. Phys.}, 13:181--192, 1959.

\bibitem{ELOP}
R.~J. Eden, P.~V. Landshoff, David~I. Olive, and J.~C. Polkinghorne.
\newblock {\em The Analytic S-Matrix}.
\newblock Cambridge University Press, 1966.

\bibitem{Nakanishi}
N.~Nakanishi.
\newblock {\em Graph Theory and {F}eynman Integrals}.
\newblock Gordon and Breach, New York, 1971.

\bibitem{Wolfram}
{Mathematica, Copyright by Wolfram Research}.

\bibitem{Hahn:2004fe}
T.~Hahn.
\newblock {CUBA: A library for multidimensional numerical integration}.
\newblock {\em Comput. Phys. Commun.}, 168:78--95, 2005.

\bibitem{Agrawal:2011tm}
S.~Agrawal, T.~Hahn, and E.~Mirabella.
\newblock {FormCalc 7}.
\newblock 2011.

\bibitem{Kawabata:1995th}
Setsuya Kawabata.
\newblock {A New version of the multidimensional integration and event
  generation package BASES/SPRING}.
\newblock {\em Comp. Phys. Commun.}, 88:309--326, 1995.

\bibitem{robodoc}
Frans Slothouber and et~al.
\newblock {ROBODoc 4.99.40}.
\newblock {\em http://www.xs4all.nl/~rfsber/Robo/robodoc.html}.

\bibitem{Bonciani:2003hc}
R.~Bonciani, P.~Mastrolia, and E.~Remiddi.
\newblock {Master integrals for the two loop QCD virtual corrections to the
  forward backward asymmetry}.
\newblock {\em Nucl.Phys.}, B690:138--176, 2004.

\bibitem{Davydychev:2003mv}
Andrei~I. Davydychev and M.Yu. Kalmykov.
\newblock {Massive Feynman diagrams and inverse binomial sums}.
\newblock {\em Nucl.Phys.}, B699:3--64, 2004.

\bibitem{Tausk:1999vh}
J.B. Tausk.
\newblock {Nonplanar massless two loop Feynman diagrams with four on-shell
  legs}.
\newblock {\em Phys.Lett.}, B469:225--234, 1999.

\bibitem{Bonciani:2008az}
R.~Bonciani, A.~Ferroglia, T.~Gehrmann, D.~Maitre, and C.~Studerus.
\newblock {Two-Loop Fermionic Corrections to Heavy-Quark Pair Production: The
  Quark-Antiquark Channel}.
\newblock {\em JHEP}, 0807:129, 2008.

\bibitem{Bonciani:2009nb}
R.~Bonciani, A.~Ferroglia, T.~Gehrmann, and C.~Studerus.
\newblock {Two-Loop Planar Corrections to Heavy-Quark Pair Production in the
  Quark-Antiquark Channel}.
\newblock {\em JHEP}, 0908:067, 2009.

\bibitem{Bonciani:2010mn}
R.~Bonciani, A.~Ferroglia, T.~Gehrmann, A.~Manteuffel, and C.~Studerus.
\newblock {Two-Loop Leading Color Corrections to Heavy-Quark Pair Production in
  the Gluon Fusion Channel}.
\newblock {\em JHEP}, 1101:102, 2011.

\bibitem{Czakon:2007ej}
M.~Czakon, A.~Mitov, and S.~Moch.
\newblock {Heavy-quark production in massless quark scattering at two loops in
  QCD}.
\newblock {\em Phys.Lett.}, B651:147--159, 2007.

\bibitem{Czakon:2007wk}
M.~Czakon, A.~Mitov, and S.~Moch.
\newblock {Heavy-quark production in gluon fusion at two loops in QCD}.
\newblock {\em Nucl.Phys.}, B798:210--250, 2008.

\bibitem{Bethke:1988zc}
S.~Bethke et~al.
\newblock {Experimental Investigation of the Energy Dependence of the Strong
  Coupling Strength}.
\newblock {\em Phys.Lett.}, B213:235, 1988.

\end{thebibliography}
